\documentclass[aps,pra,twocolumn,showpacs,amsmath,amsmath,amssymb,amsfonts]{revtex4-1}
\usepackage{graphicx}
\usepackage{dcolumn}
\usepackage{bm}
\usepackage{color}
\usepackage{multirow}
\usepackage{bbold}

\usepackage[normal]{subfigure}		% For combining several smaller figures in a single box

\newcommand{\be}{\begin{equation}}
\newcommand{\ee}{\end{equation}}
\newcommand{\ba}{\begin{align}}
\newcommand{\ea}{\end{align}}
\newcommand{\sysb}{\left\{\begin{array}}
\newcommand{\syse}{\end{array}\right.}
\newcommand{\baa}{\begin{array}}
\newcommand{\eaa}{\end{array}}
\newcommand{\bs}{\begin{split}}
\newcommand{\es}{\end{split}}

\newcommand{\matb}{\left(\begin{array}}
\newcommand{\mate}{\end{array}\right)}

\newcommand{\mal}{\mathcal}

\newcommand{\mand}{\quad\text{ and }\quad}

\newcommand{\ha}{\frac{1}{2}}

\newcommand{\lt}{\left(}
\newcommand{\rt}{\right)}
\newcommand{\lqq}{\left[}
\newcommand{\rqq}{\right]}
\newcommand{\lan}{\left\langle}
\newcommand{\ran}{\right\rangle}
\newcommand{\abs}[1]{\left| #1 \right|}
\newcommand{\eval}[1]{\left.\right|_{ #1 }}
\newcommand{\av}[1]{\lan #1 \ran}

\newcommand{\sz}{{\sigma^z}}

\newcommand{\skx}[1]{\sigma^x_{#1}}
\newcommand{\sky}[1]{\sigma^y_{#1}}
\newcommand{\skz}[1]{\sigma^z_{#1}}
\newcommand{\skp}[1]{\sigma^+_{#1}}
\newcommand{\skm}[1]{\sigma^-_{#1}}

\newcommand{\Z}{\mathbb{Z}}

\newcommand{\ket}[1]{\left| #1 \ran}
\newcommand{\bra}[1]{\lan #1 \right|}

\newcommand{\proj}[1]{\ket{#1} \bra{#1}}
\newcommand{\comm}[2]{\left[ #1, #2 \right]}
\newcommand{\acomm}[2]{\left\{ #1, #2 \right\}}

\newcommand{\reff}[1]{(\ref{#1})}

\newcommand{\suml}[2]{\sum\limits_{#1}^{#2}}

\newcommand{\comma}{\quad , \quad}

\newcommand{\gdu}{\gamma_{{\rm D}\uar}}
\newcommand{\gdd}{\gamma_{{\rm D}\dar}}
\newcommand{\gbd}{\gamma_{{\rm B}\dar}}
\newcommand{\deph}{\Gamma}
\newcommand{\ar}{\chi}

\newcommand{\dens}{\eta}

\newcommand{\dar}{\downarrow}
\newcommand{\uar}{\uparrow}

\begin{document}
\author{B. Everest, M. Marcuzzi and I. Lesanovsky}
\affiliation{School of Physics and Astronomy, The University of Nottingham, Nottingham, NG7 2RD, United Kingdom}
\title{Atomic loss and gain as a resource for non-equilibrium phase transitions\\ in optical lattices}
\date{\today}
\begin{abstract}
Recent breakthroughs in the experimental manipulation of strongly
interacting atomic Rydberg gases in lattice potentials have opened
a new avenue for the study of many-body phenomena. Considerable efforts are currently being undertaken to achieve clean experimental settings that show a minimal amount of noise and disorder and are close to zero temperature. A complementary direction investigates the interplay between coherent and dissipative processes. Recent experiments have revealed a first glimpse into the emergence of a rich non-equilibrium behavior stemming from the competition of laser excitation, strong interactions and radiative decay of Rydberg atoms. The aim of the present theoretical work is to show that local incoherent loss and gain of atoms can in fact be the source of interesting out-of-equilibrium dynamics. This perspective opens new paths for the exploration of non-equilibrium critical phenomena and, more generally, phase transitions, some of which so far have been rather difficult to study. To demonstrate the richness of the encountered dynamical behavior we consider here three examples. The first two feature local atom loss and gain together with an incoherent excitation of Rydberg states. In this setting either a continuous or a discontinuous phase transition emerges with the former being reminiscent of genuine non-equilibrium transitions of stochastic processes with multiple absorbing states. The third example considers the regime of coherent laser excitation. Here the many-body dynamics is dominated by an equilibrium transition of the ``model A'' universality class.
\end{abstract}


% 32.80.Ee 	Rydberg states

% 64.70.qj 	Dynamics and criticality

% 42.50.-p 	Quantum optics (for lasers, see 42.55.-f and 42.60.-v; see also 42.65.-k Nonlinear optics; 03.65.-w Quantum mechanics)

\maketitle
\section{Introduction}
Identifying, classifying, and understanding the emergence of collective phenomena and other many-body effects is a central objective of physics. In the past decades the refinement of experimental techniques for preparing, addressing and measuring atomic ensembles \cite{Bloch} opened entirely new possibilities for investigating not only stationary, but also dynamical properties of quantum many-body systems. Rich non-equilibrium physics often stems from the presence of competing dynamical processes and strong interactions. Amidst different platforms, gases of atoms excited to high-lying Rydberg states are currently receiving increasing attention \cite{Rydberg2, Gallagher84, Carr2013, Ebert2015, Maller2015} as they feature considerable interactions via dipole-dipole or van der Waals forces. Their interplay with the laser excitation process --- inducing coherent Rabi oscillations on the atomic populations ---  has shown to be the source of intricate behaviours, such as the formation of crystalline ground state structures \cite{Weimer08, Weimer2010, Levi2015}, collectively-enhanced Rabi oscillations \cite{Dudin2012, Barredo2014, Zeiher2015} or the emergence of correlated equilibrium states \cite{Ates2012, Ji2013, Ryd-lattice2, Garttner2013, Schempp2014}.

Since very recently, there has been considerable interest in understanding the many-body physics of interacting ensembles of Rydberg atoms in the presence of noise. While previously perceived as a detrimental feature, dissipation --- caused by fluctuating atomic level shifts or radiative decay --- can be in fact a source of an intriguingly rich dynamics. Examples include the occurence of slow or glassy dynamics \cite{PRL-KinC, Lesanovsky2014, Urvoy2015, Mattioli2015}, the relaxation into stationary states with spatial correlations \cite{AF-num1, tDMRG1, FullDynMF}, the observation of intermittency and bistabilities \cite{PRA-Int, Ryd-bistab1, FullDynMF, Experiment1} and the emergence of equilibrium \cite{MM2014, Weimer2014, Weimer2015, Experiment1, Experiment2} and out-of-equilibrium universal behavior \cite{MM_DP}.

In this work we introduce a new scenario for the study of out-of-equilibrium phases and phase transitions with Rydberg atoms. The setting we have in mind consists of a background gas --- acting as a large reservoir --- from which Rydberg states are only excited at given spatial positions which are arranged in a regular lattice. Atoms from the background dynamically enter and leave these excitation spots. In conjunction with the laser-excitation and the strong interatomic interactions this local loss and gain dynamics leads to the emergence of non-trivial many-body dynamics.

Such a scenario could be implemented in two rather different settings: in the first, a lattice of optical traps is immersed in a cold cloud of atoms and the traps are continuously filled and depleted \cite{Nogrette2014}. Current experiments aim at progressively slowing down this local dynamics
by, for example, reducing the pressure of the background gas or increasing the strength of the optical confinement \cite{Wilk2010, Barredo2015, Gaetan2009}. These attempts could be reversed and, in principle, the setup could be ``worsened'' to the point that the timescale of the loss/gain dynamics becomes comparable with the other relevant dynamical processes. The second experimental setting consists of hot atomic gases confined in thermal vapour cells. Recently it has been shown that they indeed allow the observation of correlated many-body dynamics \cite{Urvoy2015, Experiment1} when Rydberg states are excited. Our envisioned setup is then realized by restricting the laser excitation to a regular array of addressed spots. Thermal motion would move atoms in and out of these laser-illuminated regions, yielding the desired loss/gain dynamics.

\begin{figure*}
	\includegraphics[width = 0.8\textwidth]{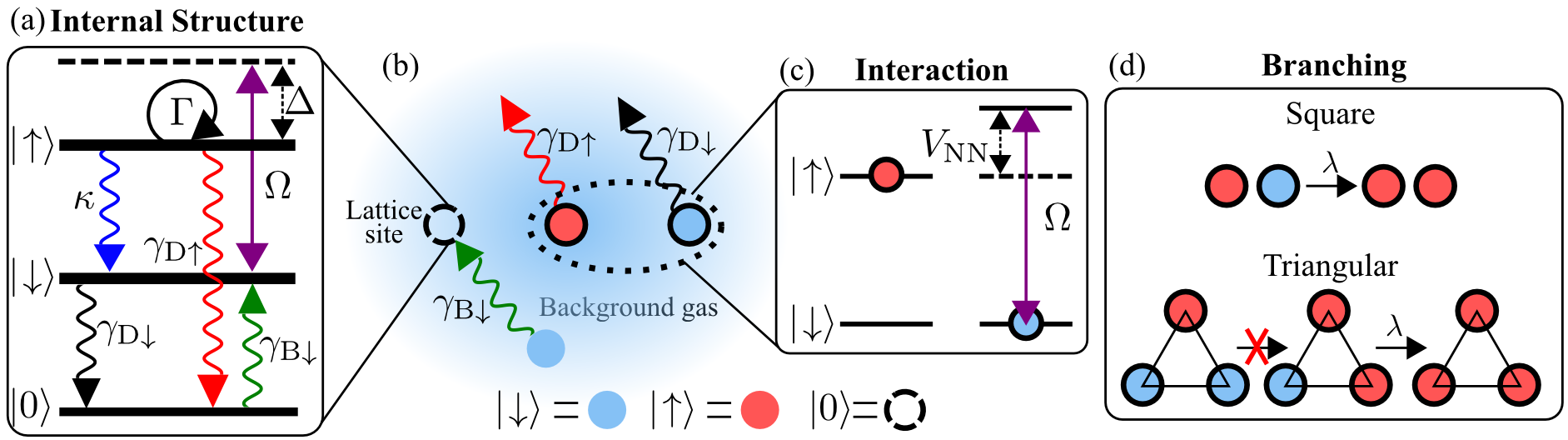}
\caption{
Schematic representation of
the system: (a,b) an optical lattice is realized within a background
cloud of atoms. Atoms within the sites undergo laser-induced coherent transitions between their ground state $\ket{\dar}$ and a high-lying (Rydberg) state $\ket{\uar}$. The corresponding Rabi frequency and laser detuning are $\Omega$ and $\Delta$, respectively. A third state, $\ket{0}$,
describes an empty or ``inactive'' site. Atoms are captured in and released from the sites with rates $\gbd$ (capturing a ground-state atom), $\gdd$ (losing a ground-state atom) and $\gdu$ (losing an excited atom). The atomic states are furthermore subject to dephasing at a rate $\deph$ and radiative decay from the Rydberg state into their ground state at rate $\kappa$. (c) Rydberg atoms interact with a van der Waals potential $V_{kq}$, whose value for nearest neighbors is denoted by $V_{\text{NN}}$. The corresponding energy shift of the Rydberg state in the vicinity of an excited atom is sketched. (d) Emergent branching processes in the limit of strong dephasing: in the case analysed in Sec.~\ref{sec:cont} (square lattice) a single excitation has the potential for branching (with rate $\lambda$) which enables the production of larger clusters. In the case discussed in Sec.~\ref{sec:first} (triangular lattice) the system parameters are chosen such that two nearby excitations are required for a cluster to grow.
}
\label{fig:f1}
\end{figure*}
Beyond introducing an additional dynamical process the consideration of local atom loss and gain might actually relax a number of challenges that are currently faced by experimentalists when studying collective many-body behavior in dissipative Rydberg lattices. It might also simplify the modelling of Rydberg gases in which typically radiative decay is accounted for as a dominant decoherence mechanism:

\begin{itemize}
 \item[(i)]It is not necessary to have (uniformly) deterministically loaded lattices, equal lattice confinement of ground state and Rydberg atoms or very low temperature states.

 \item[(ii)] It is not necessary to construct lattice potentials that equally trap ground state and Rydberg atoms. In fact it is not necessary to keep atoms trapped over an entire experimental run.

\item[(iii)]
One can employ very strongly interacting and high-lying Rydberg states that are typically long-lived. For such states the corresponding decay rate might simply be too small. In other words, it might be difficult to reach a regime in which the decay dynamics is able to properly compete with the laser excitation and interatomic interactions, which thus almost entirely characterize the evolution.

%\item[(iv)]
%\changer{
%In order to reduce the complexity of Rydberg many-body models one usually employs a few-level approximation for the atomic structure. In the simplest case one considers two levels and spontaneous emission is assumed to lead from the Rydberg level directly to the ground state. While this is certainly the dominant channel, decay often proceeds via a cascade during which several intermediate states are visited, possibly including meta-stable Rydberg levels. Depending on the overall influence of these events, one might have to account for these additional internal states, which are outside of the original model space. In our case, loss of an atom from a trap can lead to a single outcome, i.e., the trap being empty.
%}

\item[(iv)] Even when acting on timescales that set it in competition with the driving, radiative decay is inevitably accompanied by momentum kicks from photon recoil. Even when a Rydberg atom eventually decays to the desired electronic ground the resultant heating might lead to loss of the atoms which can be accounted for in our description.

\end{itemize}

For the sake of simplicity and in order to focus on the new aspects introduced by the loss/gain dynamics we will not consider radiative decay processes in this work. The underlying assumption is that the loss/gain dynamics is faster than that of the decay and/or that decay effectively induces a loss process via the mechanism described in point (v) above.

The paper is organized as follows: after defining the model in Sec.~\ref{sec:model} we introduce two scenarios that respectively feature a continuous [Sec.~\ref{sec:cont}] and discontinuous [Sec.~\ref{sec:first}] dynamical phase transition to an effective absorbing subspace of states. Here we focus on a setting where the laser excitation of Rydberg states is described by a classical rate equation \cite{MM2014-2} due to the presence of strong dephasing noise. In Sec.~\ref{sec:low-deph} we discuss the case of a coherent laser in the framework of a mean-field approach. Here we show that the dynamics is described by the so-called 'Model A' universality class, similar to what has recently been found in dissipative Rydberg gases with radiative decay \cite{Experiment2,MM2014,Weimer2014}. Concluding remarks are provided in Sec.~\ref{sec:concl}.

\section{The model}
\label{sec:model}
We employ the standard description of a Rydberg lattice gas where each atom is modeled in terms of an effective two-level system. The ground state $\ket{\dar}$ is coupled to a Rydberg $nS$-state $\ket{\uar}$ through a laser with Rabi frequency $\Omega$ and a detuning $\Delta$ with respect to the atomic transition (see Fig.~\ref{fig:f1}(a) for a visual representation). Within the rotating wave approximation the many-body Hamiltonian is then given by
\be
	H = \Omega \suml{k}{} \sigma_k^x + \Delta \suml{k}{} n_k + \ha \suml{k \neq p}{} V_{kp} n_k n_p,
	\label{eq:H}
\ee
where $V_{kp} = C_6 / \abs{\mathbf{r}_k - \mathbf{r}_p}^6$ represents the van der Waals (vdW) potential between pairs of excited atoms at positions $\mathbf{r}_k$ and $\mathbf{r}_p$, and the sum runs over all lattice sites $k = 1 \ldots L$. Interactions among ground-state atoms or between ground-state and Rydberg atoms are significantly weaker and will therefore be neglected. The operators $\sigma_k^{x/y/z}$ refer to the Pauli matrices in the $\ket{\uar}$, $\ket{\dar}$ subspace, i.e.
\be
\begin{split}
	\skx{k} & = \ket{\uar_k}\bra{\dar_k} + \ket{\dar_k} \bra{\uar_k}, \\
	 \sky{k} & = -i\ket{\uar_k}\bra{\dar_k} + i\ket{\dar_k} \bra{\uar_k}, \\
	 \skz{k} & = \ket{\uar_k}\bra{\uar_k} - \ket{\dar_k} \bra{\dar_k}.
\end{split}
\ee
Furthermore, the local density of excitations is defined as $n_k = \proj{\uar_k}$ and the density of ground state atoms as $p_k = \proj{\dar_k}$.

In order to account for atom gain/loss in the lattice sites we add an effective third state $\ket{0}$, denoting an empty (or ``inactive'') site. We also introduce the corresponding local densities of active sites $e_k = n_k + p_k =  \proj{\uar_k} + \proj{\dar_k}$.
This local loss and gain takes place with atoms from a background gas which is assumed to act as a bath. In other words, the surrounding cloud contains a much higher number of atoms than can be accommodated in the lattice and the recapture of a lost one is an unlikely event. First of all, this suppresses correlations between loss and gain processes and allows us to treat them as being independent. Secondly, since atoms are constantly exchanged with new ones no correlations can be produced in the system via these processes. Thirdly, their occurrence probabilities are not appreciably affected by the history of occupation of a given site, and can thus be considered Markovian.

The relevant processes are schematically displayed in Fig.~\ref{fig:f1} and summarized below:
\be
	 \ket{\uparrow} \,\xrightarrow{\gdu}\, \ket{0} \comma \ket{\downarrow} \,\xrightarrow{\gdd}\, \ket{0} \comma  \ket{0}  \,\xrightarrow{\gbd}\, \ket{\downarrow}.
	\label{eq:processes}
\ee
with $\gdu$, $\gdd$ and $\gbd$ being the corresponding rates. The first two processes describe the loss of a Rydberg and ground state atom, respectively, The third process corresponds to the capture of a ground state atom from the background gas. Note, that we do not consider the eventuality of Rydberg atoms being captured. The reason is that laser excitation to Rydberg states is restricted to local sites and consequently Rydberg atoms are not produced in the background gas. Hence, the transition $\ket{0} \to \ket{\uar}$ could only occur if a Rydberg atom is captured which had been previously expelled from another site, which is unlikely. Note, that we are also neglecting processes which lead to the occupation of a given site with multiple atoms. In fact, in experiments with microtraps such multi-occupancies are suppressed due to the collisional blockade \cite{Schlosser2001, Schlosser2002}. In circumstances where such suppression is not taking place the so-called dipole or Rydberg blockade \cite{Lukin2001,Urban2009,Gaetan2009} is ensuring that each site can only feature a single Rydberg excitation. This also limits the local site dynamics to a restricted space at the expense of a possibly varying local (density-dependent) Rabi frequency \cite{Cummings1983}. For the sake of simplicity we will not consider such situations in this work.

In addition to the loss/gain dynamics we consider the presence of noise, which dephases local superpositions between the states $\ket{\uar}$ and $\ket{\dar}$ at a rate $\deph$. The origin of this noise can be fluctuating background fields that result in random atomic level shifts, the broadening of atomic lines due to Doppler broadening \cite{Urvoy2015} or interaction effects \cite{Dephasing}, or a spectrally broad excitation laser \cite{Rydberg2}.

In the presence of the described coherent and dissipative processes the evolution of the density matrix $\rho$ of the system is governed by a quantum master equation in Lindblad form \cite{Lindblad76, Breuer_P}
\be
\begin{split}
	\partial_t  \rho = -i\comm{H}{\rho} + &  \suml{i,k}{} \gamma_i \lqq L_{i,k} \rho  L_{i,k}^\dag    - \ha \acomm{ L_{i,k}^\dag  L_{i,k} }{\rho}    \rqq   \\
	+ & \suml{k}{} \deph \lqq  n_k \rho n_k  -  \ha   \acomm{n_k}{\rho} \rqq .	\label{eq:QME}
\end{split}
\ee
Here the index $i$ runs over the three different sources of noise introduced in Eq.~\reff{eq:processes}, while $k$ over all lattice sites; $\acomm{A}{B} = AB + BA$ is shorthand for an anticommutator. $L_{{\rm D}\dar,k} = \ket{0_k} \bra{\dar_k}$, $L_{{\rm D}\uar,k} = \ket{0_k} \bra{\uar_k}$ and $L_{{\rm B}\dar,k} = \ket{\dar_k} \bra{0_k}$ are the corresponding local jump operators; the last term accounts for dephasing, implemented via the jump operators $n_k$.

\section{Continuous transition from/to an absorbing state}
\label{sec:cont}
We start by considering a situation in which the dephasing amplitude $\deph$ is much larger than the Rabi frequency and the other dissipative rates ($\deph \gg \Omega, \gamma_i$). In this regime, the dynamics is effectively described by means of a classical stochastic equation \cite{PRL-KinC, Ates06, Heeg2012, MM2014-2}: the underlying separation of timescales permits the adiabatic elimination of the portion of the phase space subject to dephasing. Correspondingly, the evolution of the density matrix $\rho$ of the system is projected onto the dissipation-free subspace \cite{Nakajima58, Zwanzig60, Degenfeld2014}, which in this case corresponds to the sole diagonal components in the $\sz$ basis (i.e. a basis of classical spin configurations) \cite{MM2014-2}. At the leading order in a perturbative expansion in powers of $\gamma_i / \deph$ and $\Omega / \deph$ the truncated density matrix $\mu$ evolves according to
\be
\begin{split}
	\partial_t  \mu  &   =    \suml{k}{}  \Lambda_k \lt \skx{k} \mu \skx{k}  - e_k \mu  \rt + \\
	& + \suml{i  ,k}{} \gamma_i \lqq L_{i,k} \rho  L_{i,k}^\dag    - \ha \acomm{ L_{i,k}^\dag  L_{i,k} }{\rho}    \rqq
	\label{eq:CME}
\end{split}
\ee
where
\be
	\Lambda_k = \Omega^2 \deph \lqq \lt \frac{\deph}{2} \rt^2 + \lt \Delta + \suml{q\neq k}{} V_{kq} n_q  \rt^2  \rqq^{-1}
	\label{eq:rate1}
\ee
is a configuration-dependent rate. Defining the diagonal part of the Hamiltonian ($H \eval{\Omega = 0}$) as the ``classical'' component, the second term in the brackets of Eq.~\reff{eq:rate1} corresponds to the square of the classical energy change accompanying a spin-flip at site $k$. Spin-flips that result in a significant increase or decrease in energy are therefore strongly suppressed. In the scenario investigated in this section we choose the detuning $\Delta$ such that it is opposite to the interaction energy $V_{\text{NN}}$ between neighboring excited atoms ($\Delta = -V_{\text{NN}}$) [see Fig.~\ref{fig:f1}(c)]. Hence, exciting an atom right next to an \emph{isolated} already excited one incurs no energy difference and therefore occurs at the maximal rate $\Lambda_k^{(\mathrm{max})} \equiv \lambda = 4\Omega^2 / \deph$ (see Fig.~\ref{fig:f1} (d)). We further assume to be working in a regime where $- \Delta  = V_{\text{NN}} \gg \deph$, i.e. the interaction surpasses the dephasing strength. In this regime any atom that has more than $1$ excitation in its neighborhood (or none at all), can only change its internal state at a rate of order $\Lambda_k \propto \Omega^2 \deph / \Delta^2$. This rate is significantly smaller than $\Lambda_k^{(\mathrm{max})}$ and thus such processes are strongly suppressed. For brevity, we shall refer to all of them as ``off-resonant'' processes. Among them, we emphasize that the creation of excitations in an empty neighborhood, occurring at a rate $\Lambda_k  \approx  \Omega^2 \deph / \Delta^2 $, is also strongly suppressed.

In order to gain some first insight into the expected many-body dynamics we assume for a moment that all off-resonant processes can be neglected. In this regime Eqs.~\reff{eq:CME} and \reff{eq:rate1} display features that characterize a class of stochastic processes \cite{DP_Hinrichsen}, such as the contact process and other branching-annihilating ones, which are known to undergo genuine non-equilibrium phase transitions \cite{Geza2004, NEQ_PT1}. These systems feature an absorbing phase with strictly zero density $n = \sum_k \av{n_k}/L$, i.e. in the absence of excitations no new ones can be created. Furthermore, the production of excitations can only proceed via nucleation of clusters and is, in this sense, local (it cannot occur arbitrarily far from pre-existing excitations). It is important to note that in the present case the absorbing phase does not consist of a unique state, but rather an absorbing space that is spanned by the entire set of configurations of sites which are either in state $0$ or state $\dar$. In general there is dynamics taking place within the absorbing manifold as all the absorbing configurations can be visited via the interplay of the local loss and gain processes with rates $\gdd$ and $\gbd$. This requires us to consider also the density of active sites $\dens = \sum_k \av{e_k} /L$ when analyzing the dynamics of the system.

\subsection{Mean-field approach}
The mean-field approximation discards correlations between different sites --- i.e., for every local observable $\mal{O}_k$ we substitute $\av{\mal{O}_k \mal{O}_p} \to \av{\mal{O}_k} \av{ \mal{O}_p}$ if $k \neq p$ --- and permits the formulation of closed equations of motion for the expectation values of the densities of excitations $n$ and of active sites $\dens$:
\begin{subequations}
\begin{align}
	\partial_t n &  = - \gdu  n + \suml{j=0}{z} \matb{c} z \\ j  \mate   \lambda_j   n^j \lt \dens - 2n \rt \lt 1 - n  \rt^{z-j}   \label{eq:dtn3}   \\
	\partial_t \dens &  = \gbd  -  \lt \gbd + \gdd  \rt \dens    + \lt  \gdd - \gdu   \rt n. \label{eq:dtA3}
\end{align}
\end{subequations}
Here, $z$ is the lattice coordination number (number of nearest neighbours per site) and $\lambda_j$ is shorthand for the rate of a flipping process occurring in the presence of $j$ excited neighbors. This means that $\lambda_1$ characterizes the resonant processes introduced above (and $\lambda_1 = \lambda = 4\Omega^2 / \deph$), whereas the remaining values refer to the off-resonant processes and read
\be
	\lambda_{j\neq 1} = \frac{\Omega^2 \deph}{\lt \frac{\deph}{2} \rt^2 + \Delta^2 \lt j-1 \rt^2} \approx  \frac{\Omega^2 \deph}{ \Delta^2 \lt j-1 \rt^2} \ll \lambda.
\ee
In a first approximation we neglect them altogether and Eq. (\ref{eq:dtn3}) becomes
\be
	\partial_t n   = - \gdu  n + \lambda z  n \lt \dens - 2n \rt \lt 1 - n  \rt^{z-1}.   \label{eq:dtn4}
\ee
This equation together with Eq.~\reff{eq:dtA3} predicts a transition from the region $\lambda < \lambda_c =  \gdu (\gbd + \gdd)  / (z\gbd)   $ which admits only the absorbing solution $n = 0$ to the region $\lambda > \lambda_c $ in which the system displays a finite density $n>0$ in the long-time limit. %%%%%%
%%%%%%
%%%%%%
\begin{figure}
\includegraphics[width = \columnwidth]{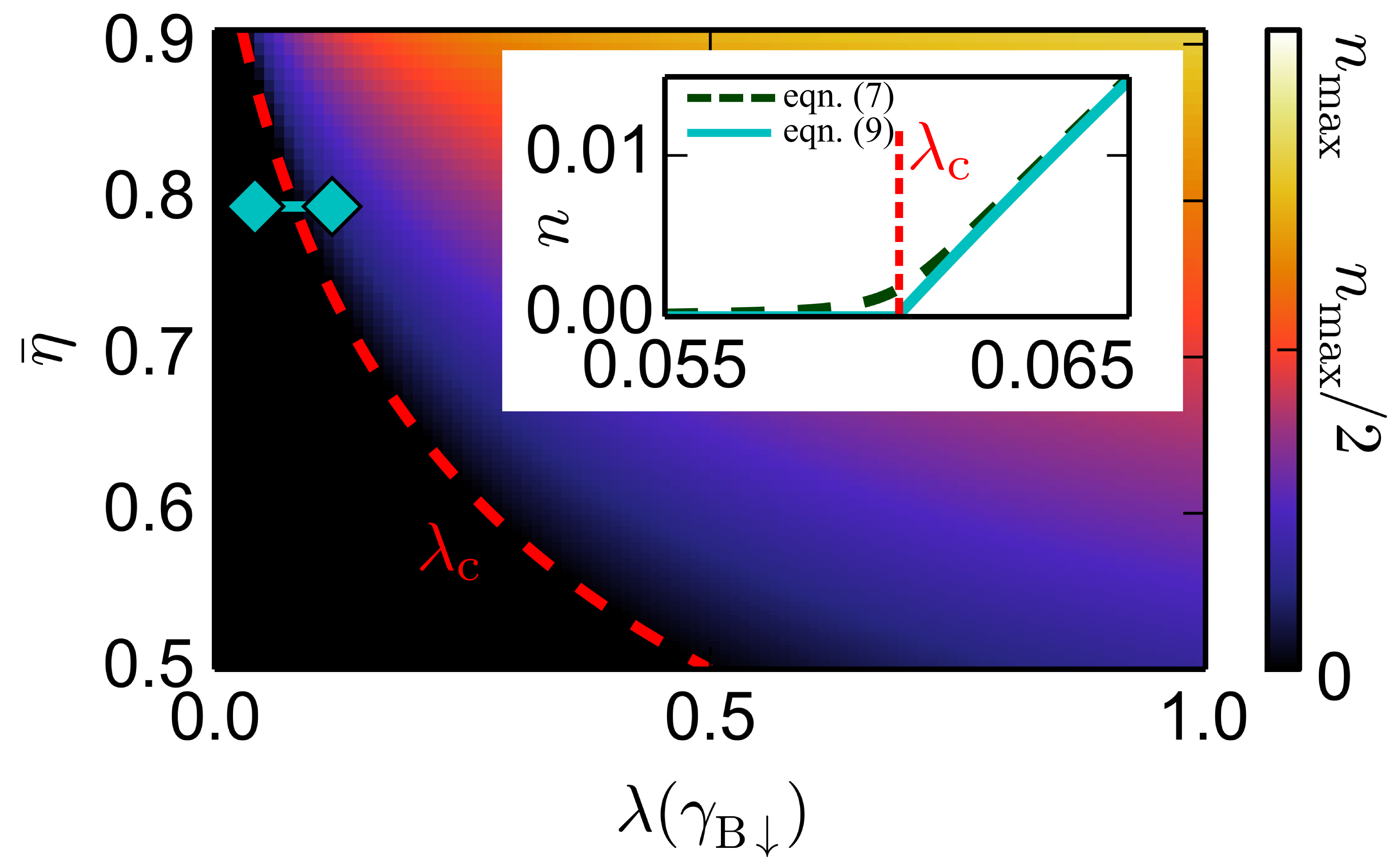}
\caption{Stationary density of excitations $n$ extracted from the mean-field equations \reff{eq:dtn4} and \reff{eq:dtA3} for $-\Delta = V_\mathrm{NN} = 64 \deph$, $\gbd = 0.01 \deph$ and $\gdu = \gdd$. The data is shown as a function of the branching rate $\lambda$ for resonant processes and the density of active sites  $\bar{\dens} = \gbd / (\gbd + \gdd) $. The color scale is bounded by $n_{\text{max}} = 0.5$. The red dashed line corresponds to the values taken by the critical rate $\lambda_c$ for different values of $\bar{\dens}$. A cross section is displayed in the inset for $\bar{\dens} = 0.8$ (along the cyan horizontal line in the main figure), which highlights the mean-field scaling behavior $n \sim \lambda - \lambda_c$. The green, dashed line corresponds to the same curve calculated including the leading off-resonant processes relevant in a Rydberg gas . As expected, the introduction of the latter makes the transition smoother, but deviations are only visible in close vicinity to the critical point $\lambda_c$.
}
\label{fig:MF_square}
\end{figure}
%%%%%%
%%%%%%
%%%%%%
In Fig.~\ref{fig:MF_square} we report the corresponding phase diagram for the choice $\Delta = - 64\deph$, $\gdu = \gdd \equiv \gamma_{\text{D}}$ in the $\lambda$--$\bar{\dens}$ plane, where the symbol $\bar{\dens}$ denotes the stationary density of active sites. The threshold value $\lambda_c$ identifies the critical point of a continuous transition between the two phases. To its right, the density scales linearly ($n \sim \lambda - \lambda_c$), while at the critical point its value decays to $0$ in time according to the power-law $n(t) \sim 1/t$. The density of active sites relaxes to the finite value $\gbd / (\gbd + \gdd)$. Consequently, at the mean-field level, this system undergoes a transition which shows some of the characteristic features of directed-percolation (DP) universality \cite{DP_Hinrichsen}.

Let us now discuss the role of the off-resonant terms. Those with $j>1$ in Eq.~\reff{eq:dtn3} do not affect the fundamental properties of the transition, as they vanish for $n \to 0$. Therefore, they can only shift the position of the critical point according to the relative statistical weights $\lambda_j$. The $j=0$ term, on the other hand, constitutes a relevant --- albeit small --- perturbation that brings the system away from the critical point. The reason is that it accounts for production of excitations in an empty neighborhood and thus prevents the aforementioned subspace of configurations devoid of excitations from being strictly absorbing. This term smooths the transition into a crossover, as highlighted in the inset of Fig.~\ref{fig:MF_square}. The magnitude of this effect can be suppressed by increasing the detuning $\Delta$. When sufficiently small it allows the observation of the mean-field scaling behavior for values of $\lambda \gtrsim \lambda_c$.

\subsection{Numerical analysis}
In order to investigate the effect of fluctuations which are not captured by the mean-field treatment we perform numerical Monte Carlo simulations, using a state in which all sites are occupied with a Rydberg atom as the initial condition. We set the rates $\gdd = \gdu \equiv \gamma_{\text{D}}$, $\gbd = 0.01 \deph$, $V_{\text{NN}} = 64 \deph = -\Delta$ and collect data for several values of the parameters $\gamma_{\text{D}}$ and $\Omega$. For this particular choice of parameters the loss and gain dynamics decouples from the excitation dynamics. This can be seen directly in Eq.~\reff{eq:dtA3} which is valid beyond mean-field. Consequently, the density of active sites $\eta$ reaches exponentially fast (on a timescale $(\gbd + \gamma_{\text{D}})^{-1}$) the steady-state value $\bar{\dens} = \gbd / (\gbd + \gamma_{\text{D}})$.

For Rydberg gases one needs to account for the fact that the off-resonant production of excitations and the long-range tails of the vdW potential affect the emergence of the phase transition. As we discuss further below, these features actually constitute a source of additional noise which to some extent may obscure the anticipated scaling behaviors. In order to shed light on the fundamental critical properties of the transition we have therefore also simulated a dynamical process in which we replace the first term of the r.h.s.~of Eq.~\reff{eq:CME} by
%
%For Rydberg gases one needs to understand in particular how the off-resonant production of excitations and the long-range tails of the vdW potential affect the emergence of the phase transition. In order to assess this in detail we have simulated --- for the sake of comparison --- a dynamical process in which we replace the first term of the r.h.s.~of Eq.~\reff{eq:CME} by
\be
	 \suml{k, i \in \left\{ k  \right\} }{} \frac{\lambda}{2d} n_i \lt  \skp{k} \mu \skm{k} - p_k \mu  \rt,
	 \label{eq:newterm}
\ee
with $\left\{ k  \right\}$ denoting the set of nearest-neighboring sites of site $k$ and $d$ being the dimension. After this replacement we have a pure branching process (as found, e.g., in the contact process mentioned above \cite{DP_Hinrichsen}) producing excitations from nearby ones at a rate $\lambda / (2d)$. The normalization by the quantity $2d$ --- which corresponds to $z$ for square lattices --- is meant to compensate for the fact that in this case multiple excitations enhance the rate.
We emphasize that, although different, the two processes we consider share the same fundamental properties: the absorbing subspace is the same and, apart from off-resonant events, branching is the only way to increase the number of excitations. Furthermore, in the presence of low-densities --- as happens in the proximity of the critical point --- the action of the branching terms in Eqs.~\reff{eq:CME} and \reff{eq:newterm} is analogous up to multiplicative factors.
For brevity, in the following we shall refer to the new stochastic process as the ``pure'' instance and to the original process [Eq.~\reff{eq:CME}] as the ``Rydberg'' one.

\begin{figure*}
\includegraphics[width = 0.8\textwidth]{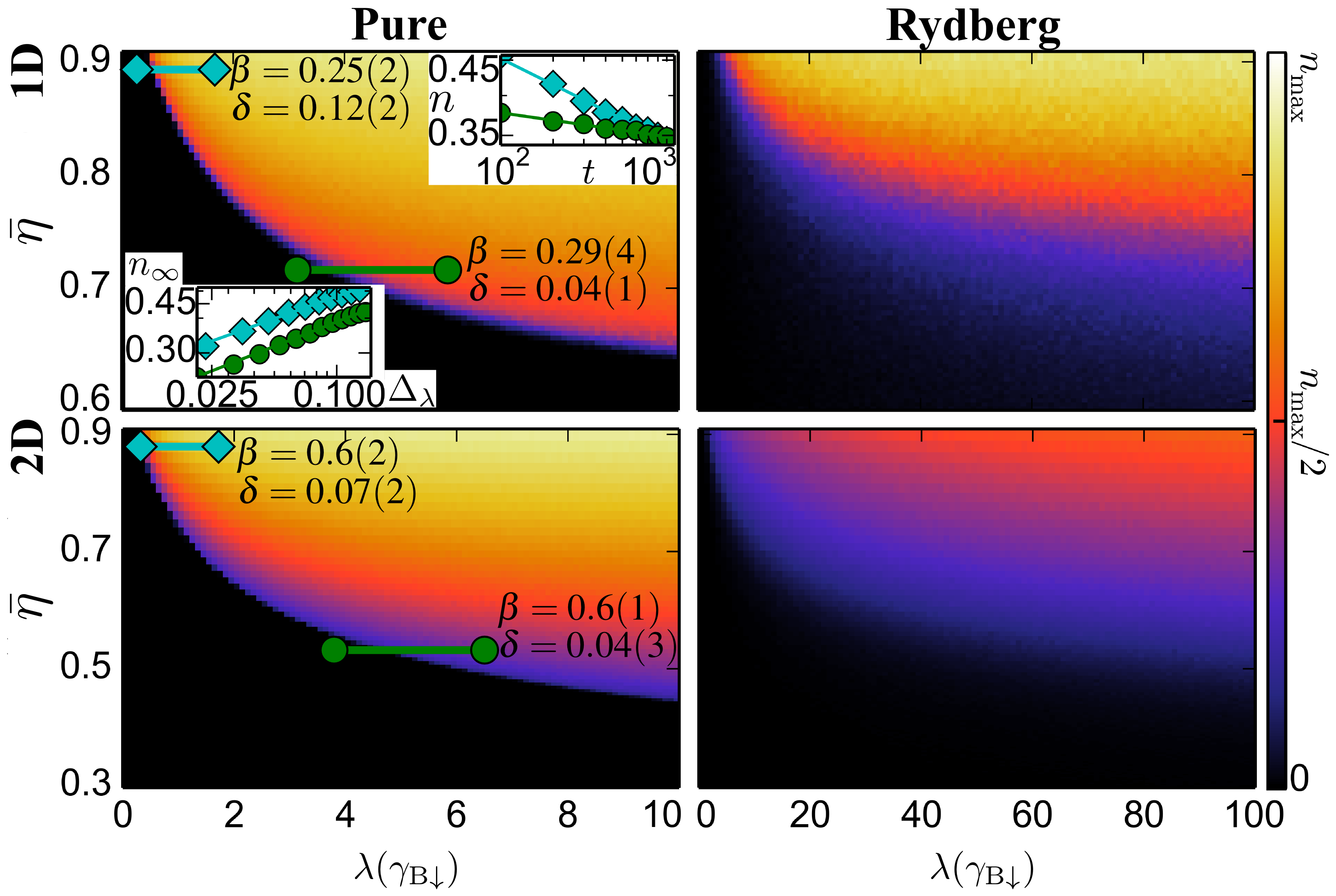}
\caption{Phase diagrams of the pure and the Rydberg process (see text) in the $\bar{\dens}$-$\lambda$ plane for a 1D chain of $100$ sites and a 2D square lattice of $20 \times 20$ sites. The parameters are chosen as $-\Delta = V_\mathrm{NN} = 64 \deph$, $\gbd = 0.01 \deph$ and $\gdd = \gdu \equiv \gamma_{\text{D}}$. The color scale is set with respect to the maximal value the density can take, i.e., $n_{\text{max}} = 1$ for the pure process and $1/2$ for the Rydberg one. Numerically computed exponents $\beta$ (static) and $\delta$ (dynamic) are displayed in the panels. The selected parameter ranges are shown as a cyan and a green line on the main plot. For the 1D pure process we also show (in log-log scale) the critical profiles of the stationary density $n$ as a function of $\Delta_\lambda = \lambda - \lambda_c$ (lower-left inset) and of its evolution in time (upper-right inset) to highlight the scaling behavior. For comparison we provide the known DP exponents \cite{DP_Hinrichsen}: $\beta_{\text{1D}} = 0.276$, $\beta_{\text{2D}} = 0.584$, $\delta_{\text{1D}} = 0.159$, and $\delta_{\text{2D}} = 0.451$.}
\label{fig:PD2D}
\end{figure*}

In Fig.~\ref{fig:PD2D} we show the stationary value (obtained from Monte Carlo simulations) of the density of excitations $n$ in the $\bar{\dens}$-$\lambda$ plane for 1D and 2D square lattices in both cases.
%%%%%%
%%%%%%
%%%%%%
The pure and the Rydberg processes display qualitatively the same behavior. As expected, in the pure case the transition from the absorbing to the active phase is significantly sharper. Beyond that, two interesting features can be observed. First, the simulations seem to suggest that the critical point $\lambda_c$ diverges
%the position of the critical point $\lambda_c$ appears to diverge
as the stationary density of active sites $\bar{\dens}$ decreases and that below a certain threshold $\bar{\dens}_c$ the transition disappears entirely. Second, there is a qualitative difference in the static and dynamic scaling behavior when varying $\bar{\dens}$. The stationary properties remain unaffected and always display, within numerical accuracy, a scaling behavior $n \sim (\lambda - \lambda_c)^\beta$ with a critical exponent $\beta$ compatible with the DP one for both one- and two-dimensional processes (see Fig.~\ref{fig:PD2D}). In contrast, the dynamical approach to stationarity changes continuously. This means that when approaching the threshold value $\bar{\dens}_c$, the critical exponent of the algebraic decay $n(t) \sim t^{-\delta}$ smoothly decreases from a value which, in 1D, is comparable with the one of pure DP to $0$.

The latter feature is strongly reminiscent of the behavior of stochastic processes with multiple absorbing states as reported in Refs. \cite{DP_Hinrichsen, Jensen1993}, which provide a qualitative explanation of our observations. Even though in our case the absorbing space is not made of individually-absorbing states, the excitation dynamics effectively perceives them as such, since it stops completely as soon as the first absorbing configuration is reached. Moreover, in the cases discussed in Refs. \cite{DP_Hinrichsen, Jensen1993} the dynamic exponent is also not constant but instead varies continuously as a function of the initial conditions, e.g. the initial density. In our simulations we start from a fixed initial condition (all atoms present and excited). However, the fast loss/gain dynamics rapidly constructs an ``effective initial condition'' with an active site density $\bar{\dens}$ determined by the rates $\gbd$ and $\gamma_{\text{D}}$. Since this initial condition is varied under a change of  $\gbd$ and $\gamma_{\text{D}}$ this might be a possible explanation for the observed variation of the dynamic exponent.

The Rydberg case features off-resonant processes and thus displays a smoothed transition. Moreover, it requires stronger driving for the active phase to appear. This can be understood by noting how clusters of excitations actually hinder their own growth. For instance, if we consider a pair of nearby excitations, elongating it to a three-excitation segment ($\uparrow\uparrow\downarrow\rightarrow\uparrow\uparrow\uparrow$) faces the presence of next-nearest-neighbor interactions. Because of them, the rate at which this process occurs is no longer $\lambda$ but instead given by
\be
	\Lambda^{(\text{NNN})}_k = \frac{\Omega^2 \deph}{ \lt \frac{\deph}{2} \rt^2 + V_{\text{NNN}}^2} .
\ee
Our choice of parameters, $V_\text{NNN} = V_\text{NN} / 2^6 = \deph$, implies $\Lambda_k^{\text{NNN}} = \lambda / 5$ and hence the branching rate is effectively reduced. Further growth along the same direction experiences much smaller corrections and thus continues approximately at a rate $\lambda / 5$. The situation worsens if we consider branching orthogonally with respect to the two original excitations, since in this case the distance between next-nearest neighbors is reduced to $\sqrt{2}$ times the lattice spacing, implying $V_{\text{NNN}} = V_{\text{NN}} / (\sqrt{2})^6 = 8\deph $ and yielding an effective rate $\Lambda_k^{\text{NNN}} = \lambda / 257$. This explains the suppression of the stationary density in the 2D case with respect to the 1D one. The relevance of this effect can be drastically reduced by partially removing the tails of the vdW interactions using a microwave dressing scheme (see Refs. \cite{Petrosyan14, Kiffner13, Kiffner13-1, MM_DP}). In other words, by coupling two Rydberg levels with a strong microwave field one can obtain a hybridization of the relative interactions. For appropriate choices, the latter features a crossover threshold separating a short-distance regime displaying the usual vdW decay from a long-distance one which is instead suppressed with respect to the previous one and can be considered approximately flat. This length-scale thus acts as a cutoff for the potential tails.

\section{Discontinuous absorbing-state phase transition}
\label{sec:first}
The continuous phase transition discussed above is not merely due to the competition between system-filling and system-emptying processes. It also strongly relies on the resonance condition requiring the presence of a single excitation nearby, which constitutes a fundamental connection with other classical branching processes \cite{DP_Hinrichsen}.

To demonstrate this we consider a 2D triangular lattice ($z=6$) of Rydberg atoms which is irradiated by a laser with detuning $-\Delta = 2 V_{\text{NN}}$. This implies that atoms now require exactly two excited neighbours to be effectively brought in resonance with the laser [see Fig.~\ref{fig:f1} (d)]. With this constraint, the geometrical structure of the lattice becomes relevant: for instance, clusters cannot grow on a square lattice, since neighboring sites do not share common neighbors.

\subsection{Mean-field approach}
\begin{figure}
	\includegraphics[width = \columnwidth]{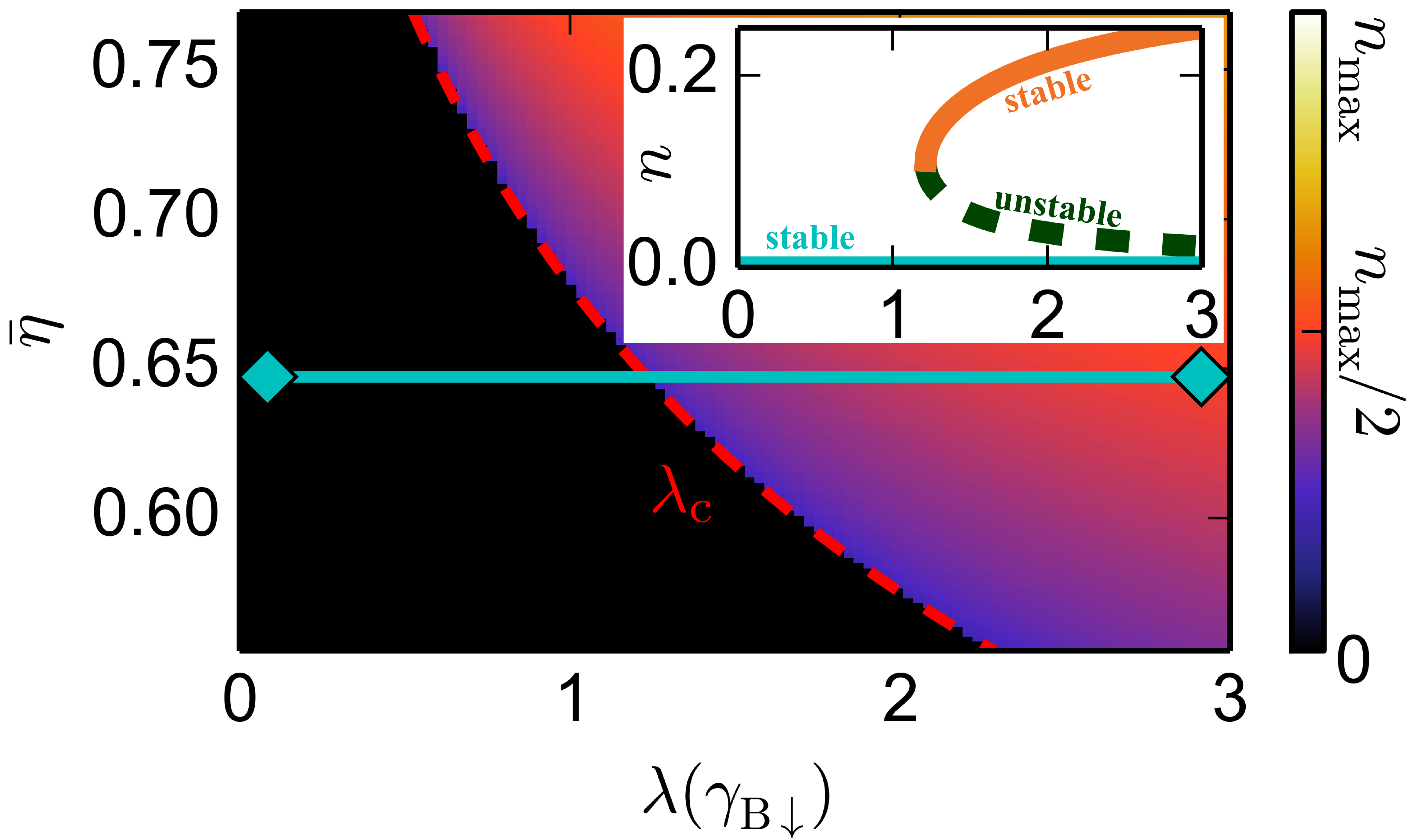}
\caption{Stationary density of excitations $n$ (we select the stable solution with highest value) extracted from the mean-field equations \reff{eq:pn} and \reff{eq:dtA3}. The data is shown as a function of $\lambda$ (branching rate) and $\bar{\dens}$ (stationary density of active sites) for $-\Delta = 2V_{\text{NN}} = 64 \deph$, $\gbd = 0.01 \deph$ and $\gdu = \gdd$. Here $n_{\text{max}} = 0.5$. The dashed line corresponds to the critical rate $\lambda_c$ for different values of $\bar{\dens}$. In the inset we plot the profiles of the three stationary solutions of Eq.~\reff{eq:pn} for $\bar{\dens} \approx 0.65$ in a parameter interval indicated by the cyan horizontal line in the main figure. The absorbing solution is displayed as a solid, cyan line which never leaves $0$. The upper solid orange line represents the stable solution with finite density which only occurs for $\lambda>\lambda_c$. The dashed green line corresponds to the third, unstable solution.
}
\label{fig:MF_tri}
\end{figure}
The density of active sites is evolving according to Eq.~\reff{eq:dtA3} in this case as well. The evolution equation for the excitation density $n$ can be cast in the form \reff{eq:dtn3}, but with $\lambda_j$ redefined as $\lambda_2 = \lambda$ and
\be
	\lambda_{j\neq 2} = \frac{\Omega^2 \deph}{\lt \frac{\deph}{2} \rt^2 + \Delta^2 \lt j-2 \rt^2} \approx  \frac{\Omega^2 \deph}{ \Delta^2 \lt j-2 \rt^2} \ll  \lambda.
\ee
Hence, the dominant contribution is now given by the $j=2$ term, i.e.,
\be
	\partial_t n   \approx - \gdu  n + \frac{2 z (z-1) \Omega^2}{\deph}  n^2 \lt \dens - 2n \rt \lt 1 - n  \rt^{z-2}.
	\label{eq:pn}
\ee
Introducing the parameter $\ar = 2z(z-1) \Omega^2 / \deph \gdu = \lambda z (z-1) / 2\gdu$, these mean-field equations predict a discontinuous transition from a phase $\ar < \ar_c$ in which the only acceptable stationary solution is $n=0$ to an active phase ($\ar > \ar_c$) in which two further solutions appear (constituting a \emph{saddle-node bifurcation} \cite{Strogatz_book}, see inset of Fig.~\ref{fig:MF_tri}). These correspond to additional real roots of the r.h.s.~of Eq.~\reff{eq:pn} and are identified by the equality $\ar n (\bar{\dens} - 2 n) = \lt 1 - n  \rt^{2-z}$. Thus, the threshold value for $\ar$ corresponds to
\be
	\ar_c = \min_{n \in \comm{0}{1/2}} \frac{1}{n \lt \bar{\dens} - 2n \rt (1-n)^{z-2}  }  .
	\label{eq:min}
\ee
Branching off from a common point (the value $n$ yielding the minimum in Eq.~\reff{eq:min}), one of these solutions is unstable (dashed green line in the inset of Fig.~\ref{fig:MF_tri}) under small perturbations and decreases asymptotically to $0$, whereas the other one (solid blue line) is stable and increases up to $\bar{\dens}/2$. The absorbing solution remains always stable. A phase diagram for this case is shown in the main panel of Fig.~\ref{fig:MF_tri}, where we display, for various choices of $\lambda$ and $\bar{\dens}$, the largest stable mean-field solution, in order to highlight the finite jump experienced by this value when the boundary $\lambda_c = 2\gdu \, \ar_c /(z (z-1))$ is crossed. The quantitative effect of the off-resonant terms is again barely noticeable at the mean-field level and we thus do not display it. We recall, however, that the subspace of configurations with $n=0$ is not perfectly absorbing and thus the lower branch is slightly lifted from $0$.
%The only qualitative difference is that the absorbing solution is slightly lifted from $n=0$ due to off-resonant production of excitations.

\subsection{Numerical analysis}
\begin{figure*}
\includegraphics[width = 0.8\textwidth]{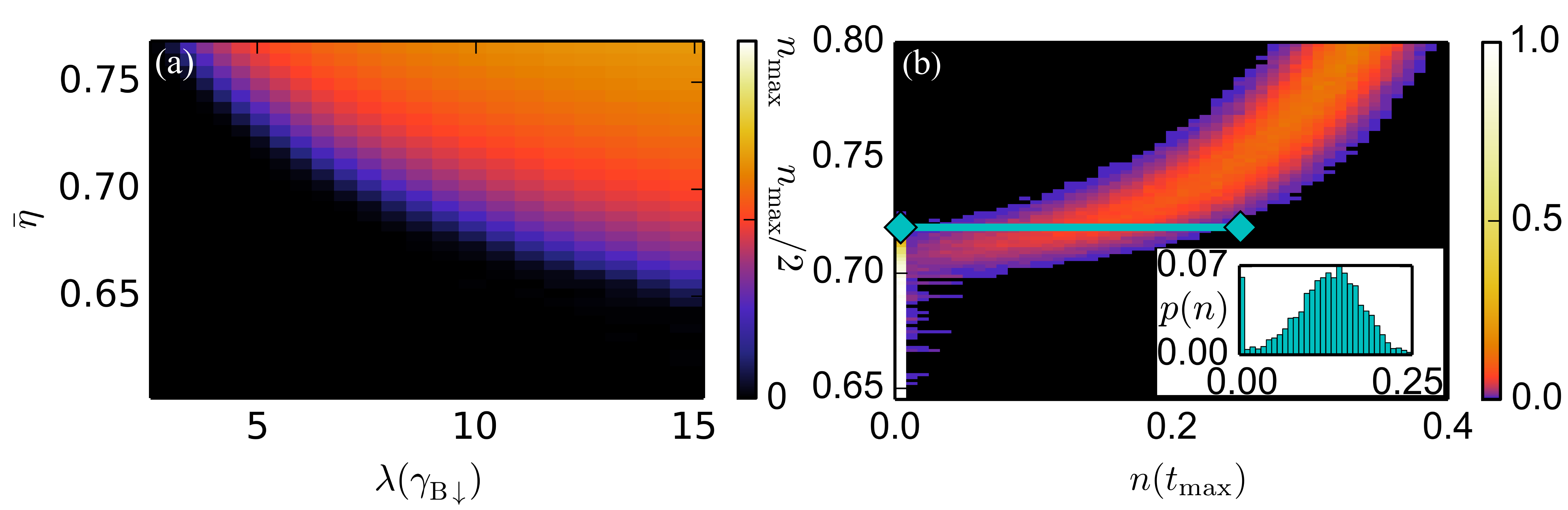}
\caption{(a) Stationary density profile in the $\lambda$-$\bar{\dens}$ plane for the branching process defined on a $25 \times 25$ triangular lattice. The color scale extends up to $n_{\text{max}} = 0.5$. As in the mean-field treatment in Fig.~\ref{fig:MF_tri}, we fix $-\Delta = 2V_{\text{NN}} = 64 \deph$, $\gbd = 0.01 \deph$ and $\gdu = \gdd \equiv \gamma_{\text{D}}$. For a sufficiently large population of trapped atoms (sufficiently small loss rate $\gamma_{\text{D}}$) and large enough $\lambda$ ($\propto \Omega^2$), the process becomes clearly capable of maintaining a finite density of excitations for long times. (b) Counting statistics $p(n)$ of the final density $n(t_{\text{max}})$ at fixed $\Omega = 0.125 \deph$ and $ \deph t_{\text{max}} = 25000 $ as a function of $\bar{\dens}$. Bimodality is a signature of the discontinuity of the phase transition and becomes clearly visible in the inset section taken along a section at $\bar{\dens} \approx 0.7194$.
}
\label{fig:tri}
\end{figure*}

To study the system beyond mean-field we employ Monte Carlo simulations. The parameters are identical to the previous simulations with the exception that $V_{\text{NN}} = -\Delta / 2 = 32 \deph $. Again, we set $\gdd = \gdu \equiv \gamma_{\text{D}}$ and use a completely filled initial state from which the fast site dynamics will generate an effectively random initial configuration. In Fig.~\ref{fig:tri} we display the stationary density $n$ in the $\Omega$-$\gamma_{\text{D}}$ plane. Note, that for the sake of clarity we do not include interactions beyond nearest neighbours, i.e. we assume that the tails of the vdW potential are modified for example by the previously mentioned microwave dressing \cite{Petrosyan14, Kiffner13, Kiffner13-1,MM_DP}. Their inclusion does not qualitatively change the phase diagram, although it makes it more difficult to highlight the discontinuous nature of the transition.
% \changer{As in that case, we checked that the phase diagram is not qualitatively modified by their inclusion.}
%%%%%%
%%%%%%
%%%%%%
%%%%%%
%%%%%%
%%%%%%
As predicted by the mean-field study we observe the presence of both an absorbing phase and an active one. At a first glance the data suggest the presence of a continuous crossover instead of a discontinuous jump. This can be reconciled with the previous Subsection's conclusions by considering that in the presence of two stable stationary solutions, different realisations of the stochastic process end up in either one or the other. However, this information is lost once the average is taken.

We have therefore computed the counting statistics --- or probability distribution --- of the excitation density $n$ at the maximal time of our simulations, $ \deph t_{\text{max}} = 15000$. It is displayed in Fig.~\ref{fig:tri}(b) and clearly highlights the discontinuous first-order nature of the transition showing that indeed, for a certain range of parameters, the distribution becomes bimodal. This confirms that some initial conditions decay into empty configurations, whereas other ones maintain a finite density $n>0$ for long times.

\section{``Model A'' universality}
\label{sec:low-deph}
For completeness we now briefly discuss the situation in which dephasing is not strong enough in order to warrant a description of the dynamics of the system in terms of effectively classical evolution equations. In this situation the full quantum master equation (\ref{eq:QME}) needs to be solved which can be done numerically only for systems consisting of a few sites. To nevertheless gain a qualitative understanding of the emerging phase structure we employ a mean-field treatment. The single-site expectation values $\av{n_k} = n$, $\av{p_k} = p$, $\av{\skx{k}} = S^x$ and $\av{\sky{k}} = S^y$ evolve according to equations
\begin{subequations}
\begin{align}
	& \dot{n} = \Omega S^y -  \gdu n    \label{eq:n}  \\[3 mm]
	& \dot{p} = -\Omega S^y  - \gdd p + \gbd (1-n-p) \\[3 mm]
	& \dot{S^x} = - (\Delta + V n) S^y - \frac{ \deph + \gdu + \gdd}{2} S^x  \label{eq:sx} \\[3 mm]
	& \dot{S^y} = 2\Omega(p-n) + (\Delta + V n) S^x -  \frac{ \deph + \gdu + \gdd}{2} S^y, \label{eq:sy}
\end{align}
\end{subequations}
where $V = \sum_{q } V_{kq} (1 - \delta_{kq})$.

First, we focus on investigating the stationary properties. Setting the time derivatives to zero one can relate all steady-state expectation values to the excitation density $n$, i.e.
\begin{subequations}
\begin{align}
		& S^y  = \frac{ \gdu}{\Omega} n    \\[3mm]
		& p  = \frac{ - \gdu n + \gbd (1-n)}{\gbd + \gdd} \\[3mm]
		& S^x   = -2\frac{\Delta + Vn}{\deph + \gdu + \gdd} \frac{ \gdu}{\Omega}n.
\end{align}
\end{subequations}
This leads a closed polynomial equation for the excitation density in the stationary state which reads
\be
	n  \lqq  a + b(c+n)^2  \rqq = 1 ,
	\label{eq:st_n}
\ee
where we have introduced the parameters
\be
\begin{split}
	a &= \frac{2\gbd + \gdu + \gdd }{ \gbd} +   \frac{( \deph + \gdu + \gdd)\gdu(\gbd + \gdd)}{4\Omega^2 \gbd}    \\[3mm]
	b &= \lt \frac{V}{\Omega} \rt^2  \frac{ \gdu(\gbd + \gdd)}{\gbd( \deph + \gdu + \gdd)}    \mand   c = \frac{\Delta}{V}.
\end{split}
\ee
Similar mean-field equations have been obtained in a number of recent works \cite{FullDynMF, Largecorr, cme1,Experiment1, MM2014} which investigate the dynamics of Rydberg gases in the presence of radiative decay.
A known feature of Eq.~\reff{eq:st_n} is that it describes a bistable behavior, not fundamentally different from the one encountered in the previous Section, i.e. for certain values of the physical parameters the mean-field equations of motion admit two, instead of one, stable stationary solutions.

\begin{figure}
	\includegraphics[width = \columnwidth]{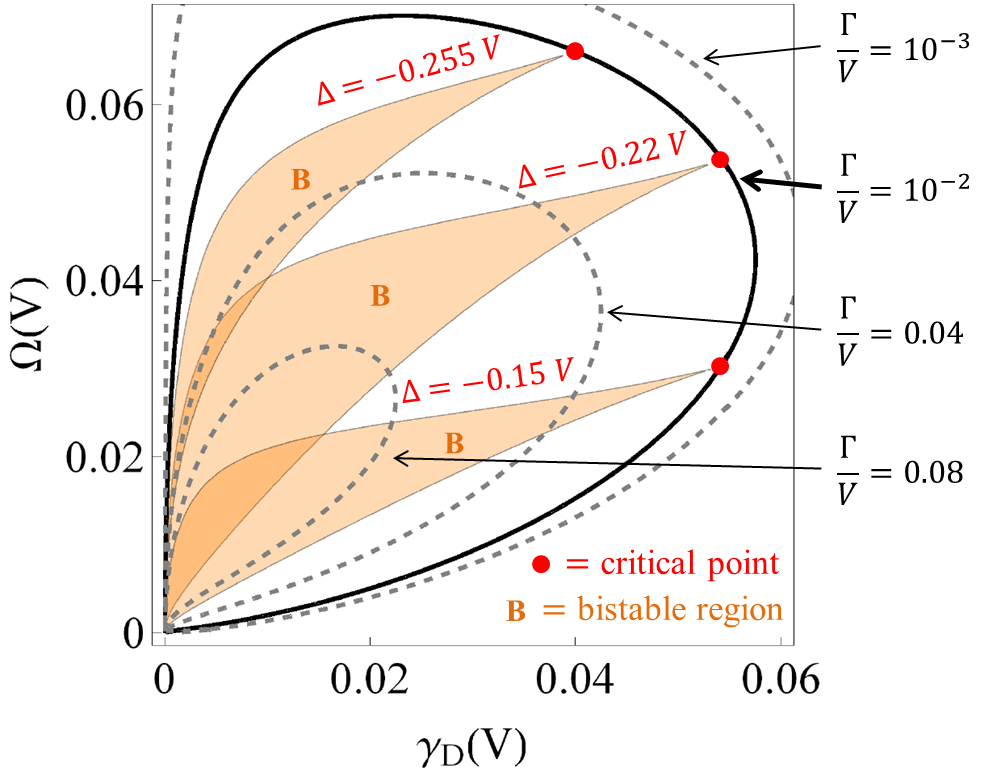}
\caption{Stationary phase diagram of the mean-field equations of motion (\ref{eq:n}-\ref{eq:sy}) for three different detunings $\Delta$ in the plane spanned by $\Omega / V$ and $\gdd / V = \gdu / V \equiv \gamma_{\text{D}} / V$. The shaded areas correspond to the bistable regions for $\deph = 0.01 V$ and $\gbd = \gamma_{\text{D}}$. Points lying outside represent choices for which there is only one physical solution. From top to bottom, the detuning is fixed at $-0.255 V$, $-0.22 V$ and $-0.15 V$. The boundaries of each shaded region meet with vanishing net angle at the corresponding critical point (red disc). The solid black line displays the path taken by the critical point as $\Delta$ is varied. The dashed lines show how the profile of this path shrinks when the dephasing rate $\deph$ is increased. From the outermost to the innermost, the values correspond to $\deph = 0.001 V$, $0.01 V$ (solid line), $0.04 V$ and $0.08 V$. The transition disappears when $\deph / V \geq \gbd / 4(\gbd + \gamma_{\text{D}}) = 0.125 $. The bistable region is also absent when $\Delta / V$ lies outside the interval $\lqq -9 \gbd/ [16 (\gbd + \gamma_{\text{D}})  ] , 0   \rqq = \lqq  -9/32, 0   \rqq$. }
\label{fig:phaseD}
\end{figure}

Figure \ref{fig:phaseD} provides a representative example of the phase diagram's structure in the $\gamma_{\text{D}}$--$\Omega$ plane, obtained in the plot for the specific choice $\gdu = \gdd = \gbd = \gamma_{\text{D}}$.
%Fixing the parameters $\gdu = \gdd = \gbd = \gamma_{\text{D}}$ leads to the phase diagram depicted in Fig.~\ref{fig:phaseD} in the $\gamma_{\text{D}}$--$\Omega$ plane. 
The bistable regions are shown as shaded areas.
%%%%%%
%%%%%%
%%%%%%
%%%%%%
%%%%%%
%%%%%%
In general, the lines delimiting these domains meet with vanishing angle in a critical point (see also the discussion in Ref.~\cite{MM2014}) which for fixed parameter $c$ is located at $a_c = -9/8c$, $b_c = -27/8c^3$. This identifies a unique direction, which in the parameter space spanned by $a$, $b$ and $c$ reads $\delta a = - \delta b \, c^2 / 9$ where $\delta a = a - a_c$ and $\delta b = b - b_c$. Varying the parameters in such a way to follow this path, the profile of the stationary density displays a branching at the critical point which can be related to the spontaneous breaking of a $\Z_2$ symmetry \cite{MM2014}. The solutions for the stationary density of excitations take here the form $n = n_c + \delta n$, with the critical density $n_c = -2c/3$ and
\be
	\delta n = \sysb{ll}  0     & (\delta b < 0)  \\
				\pm \frac{2c}{3}  \sqrt{\frac{\delta b}{\delta b + b_c}}  \sim \pm \delta b^{1/2}  & (\delta b > 0)	
	        \syse
\ee
Crossing the critical point along any other direction yields instead a behavior $\delta n \sim \delta b^{1/3}$, which relates to an explicit breaking of the symmetry. This is identical to the situation that is encountered in the thermodynamics of a ferromagnet in the presence of a finite magnetic field.

To understand the dynamical behavior near the critical point we first note that for our choice of parameters $\gdd = \gdu = \gamma_{\text{D}}$ the site dynamics decouples. The corresponding equation is
\be
	\partial_t (n+p) = \dot{\dens} = \gbd - (\gbd  + \gamma_{\text{D}}) \dens.
\ee
Linearising the remaining equations in the proximity of the critical point, we find two attractive modes and a vanishing one, signalling the presence of a critical slowing-down of the dynamics, i.e., an algebraic relaxation towards the stationary value $n(t) - n_c \sim t^{-1}$.

Recent experiments in Rydberg gases suggest that a mean-field treatment such as the one presented here can correctly capture qualitative dynamical and static features  \cite{Experiment1,Experiment2}. A recently-developed variational method \cite{Weimer2014, Weimer2015}, which improves on the mean-field, showed in this context that the bistable region is in fact 'replaced' by a first-order-transition line terminating in a critical point. The latter is related to an emerging universality belonging to the so-called ``model A'' (or Ising-Glauber) class \cite{HH}. The lines separating the bistable and stable regions can consequently be interpreted as spinodal lines, resulting in the appearance of long-lived metastable states which have been observed experimentally \cite{Experiment1} and whose properties relate to the mean-field predictions \cite{MM2014}.

Hence, the results in this section indicate that the interplay between the coherent laser excitation, the interaction and the loss/gain dynamics can drive the system towards a classical equilibrium critical point. 
Finally, we remark that here --- contrary to the cases presented in Secs. \ref{sec:cont} and \ref{sec:first}--- the presence of dephasing is inessential for the emergence of criticality. Moreover, the transition disappears when the dephasing rate exceeds 
\begin{eqnarray*}
 \deph_c = \ha \frac{\gbd}{ 2\gbd  + \gdd + \gdu } V  - \gdu -\gdd
\end{eqnarray*}
as indicated in Fig.~\ref{fig:phaseD}. From a physical point of view, this can be understood by the fact that the various cases reported here rely on facilitation effects induced by the blue-detuning of the laser with respect to the atomic transition. As discussed above, this means that the presence of excited atoms enhances the production of new ones (at a certain distance). This happens due to the fact that the interactions effectively shift the atomic levels. Therefore, if the broadening of the latter induced by dissipation becomes comparable with the shift itself (e.g., $\deph \approx V$), the facilitation effect disappears and so does the critical behavior.

\section{Conclusions}
\label{sec:concl}
In this work we have illustrated a number of non-equilibrium phenomena that can be explored within lattice gases of Rydberg atoms in the presence of local loss and gain processes. As a matter of fact, the latter, which are currently regarded as unwanted sources of noise to overcome, turned out to be key for the emergence of the collective dynamics reported above.

In the limit of strong dephasing the Rydberg gas realizes several instances of transitions from and to absorbing states. Facilitating the flipping of atoms which possess a single excited neighbor yields a continuous transition whose dynamic properties vary smoothly depending on the average density of active sites, while the static ones remain unchanged. This is strongly reminiscent of the behavior of stochastic processes with multiple absorbing states. On the other hand, on a triangular lattice and with a resonance condition asking for two neighbors to be excited the system undergoes a discontinuous phase transition, highlighted by the bimodal structure of the stationary density of excitations.

In the limit where the coherent laser excitation dominates the dephasing processes the static and dynamical properties of the system are determined by an equilibrium critical point belonging to the ``model A'' universality class. This hints at the possibility of investigating these collective behaviors in a setting complementary to the ones in which these phenomena were previously studied, all of which relied on the presence of radiative decay.

A particularly appealing aspect of the present study is the prospect to explore non-equilibrium phase transitions with absorbing states. Although the underlying universality classes -- all of which relate, to a certain degree, to directed percolation --- have been thoroughly investigated in the past, there are not many condensed matter systems which are known to display the corresponding critical behaviors \cite{Hinrichsen_exp, DP_exp, DP_exp_long}. Rydberg gases in the presence of local loss and gain permit the detailed exploration of directed percolation universality in all three dimensions \cite{MM_DP} and also in the presence of a manifold of absorbing states. Peculiar features such as the dependence of the critical dynamics on the initial state, which have been studied in the framework of idealized model systems, can now be in principle explored experimentally.

\begin{acknowledgments}
The authors would like to acknowledge R. Gutierrez and E. Levi for useful discussions. We are thankful for access to the University of Nottingham High Performance Computing Facility. The research leading to these results has received funding from the European Research Council under the European Union's Seventh Framework Programme (FP/2007-2013) / ERC Grant Agreement n. 335266 (ESCQUMA). Further funding was received through the H2020-FETPROACT-2014 grant No.  640378 (RYSQ). We also acknowledge financial support from EPSRC Grant no.\ EP/J009776/1.
\end{acknowledgments}

\bibliography{biblio}

\begin{thebibliography}{64}%
\makeatletter
\providecommand \@ifxundefined [1]{%
 \@ifx{#1\undefined}
}%
\providecommand \@ifnum [1]{%
 \ifnum #1\expandafter \@firstoftwo
 \else \expandafter \@secondoftwo
 \fi
}%
\providecommand \@ifx [1]{%
 \ifx #1\expandafter \@firstoftwo
 \else \expandafter \@secondoftwo
 \fi
}%
\providecommand \natexlab [1]{#1}%
\providecommand \enquote  [1]{``#1''}%
\providecommand \bibnamefont  [1]{#1}%
\providecommand \bibfnamefont [1]{#1}%
\providecommand \citenamefont [1]{#1}%
\providecommand \href@noop [0]{\@secondoftwo}%
\providecommand \href [0]{\begingroup \@sanitize@url \@href}%
\providecommand \@href[1]{\@@startlink{#1}\@@href}%
\providecommand \@@href[1]{\endgroup#1\@@endlink}%
\providecommand \@sanitize@url [0]{\catcode `\\12\catcode `\$12\catcode
  `\&12\catcode `\#12\catcode `\^12\catcode `\_12\catcode `\%12\relax}%
\providecommand \@@startlink[1]{}%
\providecommand \@@endlink[0]{}%
\providecommand \url  [0]{\begingroup\@sanitize@url \@url }%
\providecommand \@url [1]{\endgroup\@href {#1}{\urlprefix }}%
\providecommand \urlprefix  [0]{URL }%
\providecommand \Eprint [0]{\href }%
\providecommand \doibase [0]{http://dx.doi.org/}%
\providecommand \selectlanguage [0]{\@gobble}%
\providecommand \bibinfo  [0]{\@secondoftwo}%
\providecommand \bibfield  [0]{\@secondoftwo}%
\providecommand \translation [1]{[#1]}%
\providecommand \BibitemOpen [0]{}%
\providecommand \bibitemStop [0]{}%
\providecommand \bibitemNoStop [0]{.\EOS\space}%
\providecommand \EOS [0]{\spacefactor3000\relax}%
\providecommand \BibitemShut  [1]{\csname bibitem#1\endcsname}%
\let\auto@bib@innerbib\@empty
%</preamble>
\bibitem [{\citenamefont {Bloch}\ \emph {et~al.}(2008)\citenamefont {Bloch},
  \citenamefont {Dalibard},\ and\ \citenamefont {Zwerger}}]{Bloch}%
  \BibitemOpen
  \bibfield  {author} {\bibinfo {author} {\bibfnamefont {I.}~\bibnamefont
  {Bloch}}, \bibinfo {author} {\bibfnamefont {J.}~\bibnamefont {Dalibard}}, \
  and\ \bibinfo {author} {\bibfnamefont {W.}~\bibnamefont {Zwerger}},\
  }\href@noop {} {\bibfield  {journal} {\bibinfo  {journal} {Rev. Mod. Phys.}\
  }\textbf {\bibinfo {volume} {80}},\ \bibinfo {pages} {885} (\bibinfo {year}
  {2008})}\BibitemShut {NoStop}%
\bibitem [{\citenamefont {{L\"ow}}\ \emph {et~al.}(2012)\citenamefont
  {{L\"ow}}, \citenamefont {Weimer}, \citenamefont {Nipper}, \citenamefont
  {Balewski}, \citenamefont {Butscher}, \citenamefont {{B\"uchler}},\ and\
  \citenamefont {Pfau}}]{Rydberg2}%
  \BibitemOpen
  \bibfield  {author} {\bibinfo {author} {\bibfnamefont {R.}~\bibnamefont
  {{L\"ow}}}, \bibinfo {author} {\bibfnamefont {H.}~\bibnamefont {Weimer}},
  \bibinfo {author} {\bibfnamefont {J.}~\bibnamefont {Nipper}}, \bibinfo
  {author} {\bibfnamefont {J.~B.}\ \bibnamefont {Balewski}}, \bibinfo {author}
  {\bibfnamefont {B.}~\bibnamefont {Butscher}}, \bibinfo {author}
  {\bibfnamefont {H.~P.}\ \bibnamefont {{B\"uchler}}}, \ and\ \bibinfo {author}
  {\bibfnamefont {T.}~\bibnamefont {Pfau}},\ }\href@noop {} {\bibfield
  {journal} {\bibinfo  {journal} {J. Phys. B: At. Mol. Opt. Phys.}\ }\textbf
  {\bibinfo {volume} {45}},\ \bibinfo {pages} {113001} (\bibinfo {year}
  {2012})}\BibitemShut {NoStop}%
\bibitem [{\citenamefont {Gallagher}(1984)}]{Gallagher84}%
  \BibitemOpen
  \bibfield  {author} {\bibinfo {author} {\bibfnamefont {T.}~\bibnamefont
  {Gallagher}},\ }\href@noop {} {\emph {\bibinfo {title} {Rydberg Atoms}}}\
  (\bibinfo  {publisher} {Cambridge University Press},\ \bibinfo {year}
  {1984})\BibitemShut {NoStop}%
\bibitem [{\citenamefont {Carr}\ and\ \citenamefont
  {Saffman}(2013)}]{Carr2013}%
  \BibitemOpen
  \bibfield  {author} {\bibinfo {author} {\bibfnamefont {A.~W.}\ \bibnamefont
  {Carr}}\ and\ \bibinfo {author} {\bibfnamefont {M.}~\bibnamefont {Saffman}},\
  }\href {\doibase 10.1103/PhysRevLett.111.033607} {\bibfield  {journal}
  {\bibinfo  {journal} {Phys. Rev. Lett.}\ }\textbf {\bibinfo {volume} {111}},\
  \bibinfo {pages} {033607} (\bibinfo {year} {2013})}\BibitemShut {NoStop}%
\bibitem [{\citenamefont {{Ebert}}\ \emph {et~al.}(2015)\citenamefont
  {{Ebert}}, \citenamefont {{Kwon}}, \citenamefont {{Walker}},\ and\
  \citenamefont {{Saffman}}}]{Ebert2015}%
  \BibitemOpen
  \bibfield  {author} {\bibinfo {author} {\bibfnamefont {M.}~\bibnamefont
  {{Ebert}}}, \bibinfo {author} {\bibfnamefont {M.}~\bibnamefont {{Kwon}}},
  \bibinfo {author} {\bibfnamefont {T.~G.}\ \bibnamefont {{Walker}}}, \ and\
  \bibinfo {author} {\bibfnamefont {M.}~\bibnamefont {{Saffman}}},\ }\href@noop
  {} {\bibfield  {journal} {\bibinfo  {journal} {ArXiv e-prints}\ } (\bibinfo
  {year} {2015})},\ \Eprint {http://arxiv.org/abs/1501.04083} {arXiv:1501.04083
  [quant-ph]} \BibitemShut {NoStop}%
\bibitem [{\citenamefont {{Maller}}\ \emph {et~al.}(2015)\citenamefont
  {{Maller}}, \citenamefont {{Lichtman}}, \citenamefont {{Xia}}, \citenamefont
  {{Sun}}, \citenamefont {{Piotrowicz}}, \citenamefont {{Carr}}, \citenamefont
  {{Isenhower}},\ and\ \citenamefont {{Saffman}}}]{Maller2015}%
  \BibitemOpen
  \bibfield  {author} {\bibinfo {author} {\bibfnamefont {K.~M.}\ \bibnamefont
  {{Maller}}}, \bibinfo {author} {\bibfnamefont {M.~T.}\ \bibnamefont
  {{Lichtman}}}, \bibinfo {author} {\bibfnamefont {T.}~\bibnamefont {{Xia}}},
  \bibinfo {author} {\bibfnamefont {Y.}~\bibnamefont {{Sun}}}, \bibinfo
  {author} {\bibfnamefont {M.~J.}\ \bibnamefont {{Piotrowicz}}}, \bibinfo
  {author} {\bibfnamefont {A.~W.}\ \bibnamefont {{Carr}}}, \bibinfo {author}
  {\bibfnamefont {L.}~\bibnamefont {{Isenhower}}}, \ and\ \bibinfo {author}
  {\bibfnamefont {M.}~\bibnamefont {{Saffman}}},\ }\href@noop {} {\bibfield
  {journal} {\bibinfo  {journal} {ArXiv e-prints}\ } (\bibinfo {year}
  {2015})},\ \Eprint {http://arxiv.org/abs/1506.06416} {arXiv:1506.06416
  [quant-ph]} \BibitemShut {NoStop}%
\bibitem [{\citenamefont {Weimer}\ \emph {et~al.}(2008)\citenamefont {Weimer},
  \citenamefont {L\"{o}w}, \citenamefont {Pfau},\ and\ \citenamefont
  {B\"{u}chler}}]{Weimer08}%
  \BibitemOpen
  \bibfield  {author} {\bibinfo {author} {\bibfnamefont {H.}~\bibnamefont
  {Weimer}}, \bibinfo {author} {\bibfnamefont {R.}~\bibnamefont {L\"{o}w}},
  \bibinfo {author} {\bibfnamefont {T.}~\bibnamefont {Pfau}}, \ and\ \bibinfo
  {author} {\bibfnamefont {H.~P.}\ \bibnamefont {B\"{u}chler}},\ }\href@noop {}
  {\bibfield  {journal} {\bibinfo  {journal} {Phys. Rev. Lett.}\ }\textbf
  {\bibinfo {volume} {101}},\ \bibinfo {pages} {250601} (\bibinfo {year}
  {2008})}\BibitemShut {NoStop}%
\bibitem [{\citenamefont {Weimer}\ and\ \citenamefont
  {B\"uchler}(2010)}]{Weimer2010}%
  \BibitemOpen
  \bibfield  {author} {\bibinfo {author} {\bibfnamefont {H.}~\bibnamefont
  {Weimer}}\ and\ \bibinfo {author} {\bibfnamefont {H.~P.}\ \bibnamefont
  {B\"uchler}},\ }\href {\doibase 10.1103/PhysRevLett.105.230403} {\bibfield
  {journal} {\bibinfo  {journal} {Phys. Rev. Lett.}\ }\textbf {\bibinfo
  {volume} {105}},\ \bibinfo {pages} {230403} (\bibinfo {year}
  {2010})}\BibitemShut {NoStop}%
\bibitem [{\citenamefont {{Levi}}\ \emph {et~al.}(2015)\citenamefont {{Levi}},
  \citenamefont {{Min{\'a}{\v r}}}, \citenamefont {{Garrahan}},\ and\
  \citenamefont {{Lesanovsky}}}]{Levi2015}%
  \BibitemOpen
  \bibfield  {author} {\bibinfo {author} {\bibfnamefont {E.}~\bibnamefont
  {{Levi}}}, \bibinfo {author} {\bibfnamefont {J.}~\bibnamefont {{Min{\'a}{\v
  r}}}}, \bibinfo {author} {\bibfnamefont {J.~P.}\ \bibnamefont {{Garrahan}}},
  \ and\ \bibinfo {author} {\bibfnamefont {I.}~\bibnamefont {{Lesanovsky}}},\
  }\href@noop {} {\bibfield  {journal} {\bibinfo  {journal} {ArXiv e-prints}\ }
  (\bibinfo {year} {2015})},\ \Eprint {http://arxiv.org/abs/1503.03259}
  {arXiv:1503.03259 [physics.atom-ph]} \BibitemShut {NoStop}%
\bibitem [{\citenamefont {Dudin}\ \emph {et~al.}(2012)\citenamefont {Dudin},
  \citenamefont {Li}, \citenamefont {Bariani},\ and\ \citenamefont
  {Kuzmich}}]{Dudin2012}%
  \BibitemOpen
  \bibfield  {author} {\bibinfo {author} {\bibfnamefont {Y.}~\bibnamefont
  {Dudin}}, \bibinfo {author} {\bibfnamefont {L.}~\bibnamefont {Li}}, \bibinfo
  {author} {\bibfnamefont {F.}~\bibnamefont {Bariani}}, \ and\ \bibinfo
  {author} {\bibfnamefont {A.}~\bibnamefont {Kuzmich}},\ }\href@noop {}
  {\bibfield  {journal} {\bibinfo  {journal} {Nature Physics}\ }\textbf
  {\bibinfo {volume} {8}},\ \bibinfo {pages} {790} (\bibinfo {year}
  {2012})}\BibitemShut {NoStop}%
\bibitem [{\citenamefont {Barredo}\ \emph {et~al.}(2014)\citenamefont
  {Barredo}, \citenamefont {Ravets}, \citenamefont {Labuhn}, \citenamefont
  {B\'eguin}, \citenamefont {Vernier}, \citenamefont {Nogrette}, \citenamefont
  {Lahaye},\ and\ \citenamefont {Browaeys}}]{Barredo2014}%
  \BibitemOpen
  \bibfield  {author} {\bibinfo {author} {\bibfnamefont {D.}~\bibnamefont
  {Barredo}}, \bibinfo {author} {\bibfnamefont {S.}~\bibnamefont {Ravets}},
  \bibinfo {author} {\bibfnamefont {H.}~\bibnamefont {Labuhn}}, \bibinfo
  {author} {\bibfnamefont {L.}~\bibnamefont {B\'eguin}}, \bibinfo {author}
  {\bibfnamefont {A.}~\bibnamefont {Vernier}}, \bibinfo {author} {\bibfnamefont
  {F.}~\bibnamefont {Nogrette}}, \bibinfo {author} {\bibfnamefont
  {T.}~\bibnamefont {Lahaye}}, \ and\ \bibinfo {author} {\bibfnamefont
  {A.}~\bibnamefont {Browaeys}},\ }\href {\doibase
  10.1103/PhysRevLett.112.183002} {\bibfield  {journal} {\bibinfo  {journal}
  {Phys. Rev. Lett.}\ }\textbf {\bibinfo {volume} {112}},\ \bibinfo {pages}
  {183002} (\bibinfo {year} {2014})}\BibitemShut {NoStop}%
\bibitem [{\citenamefont {{Zeiher}}\ \emph {et~al.}(2015)\citenamefont
  {{Zeiher}}, \citenamefont {{Schau{\ss}}}, \citenamefont {{Hild}},
  \citenamefont {{Macr{\`i}}}, \citenamefont {{Bloch}},\ and\ \citenamefont
  {{Gross}}}]{Zeiher2015}%
  \BibitemOpen
  \bibfield  {author} {\bibinfo {author} {\bibfnamefont {J.}~\bibnamefont
  {{Zeiher}}}, \bibinfo {author} {\bibfnamefont {P.}~\bibnamefont
  {{Schau{\ss}}}}, \bibinfo {author} {\bibfnamefont {S.}~\bibnamefont
  {{Hild}}}, \bibinfo {author} {\bibfnamefont {T.}~\bibnamefont {{Macr{\`i}}}},
  \bibinfo {author} {\bibfnamefont {I.}~\bibnamefont {{Bloch}}}, \ and\
  \bibinfo {author} {\bibfnamefont {C.}~\bibnamefont {{Gross}}},\ }\href@noop
  {} {\bibfield  {journal} {\bibinfo  {journal} {ArXiv e-prints}\ } (\bibinfo
  {year} {2015})},\ \Eprint {http://arxiv.org/abs/1503.02452} {arXiv:1503.02452
  [physics.atom-ph]} \BibitemShut {NoStop}%
\bibitem [{\citenamefont {Ates}\ \emph
  {et~al.}(2012{\natexlab{a}})\citenamefont {Ates}, \citenamefont {Garrahan},\
  and\ \citenamefont {Lesanovsky}}]{Ates2012}%
  \BibitemOpen
  \bibfield  {author} {\bibinfo {author} {\bibfnamefont {C.}~\bibnamefont
  {Ates}}, \bibinfo {author} {\bibfnamefont {J.~P.}\ \bibnamefont {Garrahan}},
  \ and\ \bibinfo {author} {\bibfnamefont {I.}~\bibnamefont {Lesanovsky}},\
  }\href {\doibase 10.1103/PhysRevLett.108.110603} {\bibfield  {journal}
  {\bibinfo  {journal} {Phys. Rev. Lett.}\ }\textbf {\bibinfo {volume} {108}},\
  \bibinfo {pages} {110603} (\bibinfo {year} {2012}{\natexlab{a}})}\BibitemShut
  {NoStop}%
\bibitem [{\citenamefont {Ji}\ \emph {et~al.}(2013)\citenamefont {Ji},
  \citenamefont {Ates}, \citenamefont {Garrahan},\ and\ \citenamefont
  {Lesanovsky}}]{Ji2013}%
  \BibitemOpen
  \bibfield  {author} {\bibinfo {author} {\bibfnamefont {S.}~\bibnamefont
  {Ji}}, \bibinfo {author} {\bibfnamefont {C.}~\bibnamefont {Ates}}, \bibinfo
  {author} {\bibfnamefont {J.~P.}\ \bibnamefont {Garrahan}}, \ and\ \bibinfo
  {author} {\bibfnamefont {I.}~\bibnamefont {Lesanovsky}},\ }\href
  {http://stacks.iop.org/1742-5468/2013/i=02/a=P02005} {\bibfield  {journal}
  {\bibinfo  {journal} {Journal of Statistical Mechanics: Theory and
  Experiment}\ }\textbf {\bibinfo {volume} {2013}},\ \bibinfo {pages} {P02005}
  (\bibinfo {year} {2013})}\BibitemShut {NoStop}%
\bibitem [{\citenamefont {Schau{\ss}}\ \emph {et~al.}(2012)\citenamefont
  {Schau{\ss}}, \citenamefont {Cheneau}, \citenamefont {Endres}, \citenamefont
  {Fukuhara}, \citenamefont {Hild}, \citenamefont {Omran}, \citenamefont
  {Pohl}, \citenamefont {Gross}, \citenamefont {Kuhr},\ and\ \citenamefont
  {Bloch}}]{Ryd-lattice2}%
  \BibitemOpen
  \bibfield  {author} {\bibinfo {author} {\bibfnamefont {P.}~\bibnamefont
  {Schau{\ss}}}, \bibinfo {author} {\bibfnamefont {M.}~\bibnamefont {Cheneau}},
  \bibinfo {author} {\bibfnamefont {M.}~\bibnamefont {Endres}}, \bibinfo
  {author} {\bibfnamefont {T.}~\bibnamefont {Fukuhara}}, \bibinfo {author}
  {\bibfnamefont {S.}~\bibnamefont {Hild}}, \bibinfo {author} {\bibfnamefont
  {A.}~\bibnamefont {Omran}}, \bibinfo {author} {\bibfnamefont
  {T.}~\bibnamefont {Pohl}}, \bibinfo {author} {\bibfnamefont {C.}~\bibnamefont
  {Gross}}, \bibinfo {author} {\bibfnamefont {S.}~\bibnamefont {Kuhr}}, \ and\
  \bibinfo {author} {\bibfnamefont {I.}~\bibnamefont {Bloch}},\ }\href@noop {}
  {\bibfield  {journal} {\bibinfo  {journal} {Nature}\ }\textbf {\bibinfo
  {volume} {491}},\ \bibinfo {pages} {87} (\bibinfo {year} {2012})}\BibitemShut
  {NoStop}%
\bibitem [{\citenamefont {G\"arttner}\ \emph {et~al.}(2013)\citenamefont
  {G\"arttner}, \citenamefont {Heeg}, \citenamefont {Gasenzer},\ and\
  \citenamefont {Evers}}]{Garttner2013}%
  \BibitemOpen
  \bibfield  {author} {\bibinfo {author} {\bibfnamefont {M.}~\bibnamefont
  {G\"arttner}}, \bibinfo {author} {\bibfnamefont {K.~P.}\ \bibnamefont
  {Heeg}}, \bibinfo {author} {\bibfnamefont {T.}~\bibnamefont {Gasenzer}}, \
  and\ \bibinfo {author} {\bibfnamefont {J.}~\bibnamefont {Evers}},\ }\href
  {\doibase 10.1103/PhysRevA.88.043410} {\bibfield  {journal} {\bibinfo
  {journal} {Phys. Rev. A}\ }\textbf {\bibinfo {volume} {88}},\ \bibinfo
  {pages} {043410} (\bibinfo {year} {2013})}\BibitemShut {NoStop}%
\bibitem [{\citenamefont {Schempp}\ \emph {et~al.}(2014)\citenamefont
  {Schempp}, \citenamefont {G\"unter}, \citenamefont {Robert-de Saint-Vincent},
  \citenamefont {Hofmann}, \citenamefont {Breyel}, \citenamefont {Komnik},
  \citenamefont {Sch\"onleber}, \citenamefont {G\"arttner}, \citenamefont
  {Evers}, \citenamefont {Whitlock},\ and\ \citenamefont
  {Weidem\"uller}}]{Schempp2014}%
  \BibitemOpen
  \bibfield  {author} {\bibinfo {author} {\bibfnamefont {H.}~\bibnamefont
  {Schempp}}, \bibinfo {author} {\bibfnamefont {G.}~\bibnamefont {G\"unter}},
  \bibinfo {author} {\bibfnamefont {M.}~\bibnamefont {Robert-de
  Saint-Vincent}}, \bibinfo {author} {\bibfnamefont {C.~S.}\ \bibnamefont
  {Hofmann}}, \bibinfo {author} {\bibfnamefont {D.}~\bibnamefont {Breyel}},
  \bibinfo {author} {\bibfnamefont {A.}~\bibnamefont {Komnik}}, \bibinfo
  {author} {\bibfnamefont {D.~W.}\ \bibnamefont {Sch\"onleber}}, \bibinfo
  {author} {\bibfnamefont {M.}~\bibnamefont {G\"arttner}}, \bibinfo {author}
  {\bibfnamefont {J.}~\bibnamefont {Evers}}, \bibinfo {author} {\bibfnamefont
  {S.}~\bibnamefont {Whitlock}}, \ and\ \bibinfo {author} {\bibfnamefont
  {M.}~\bibnamefont {Weidem\"uller}},\ }\href {\doibase
  10.1103/PhysRevLett.112.013002} {\bibfield  {journal} {\bibinfo  {journal}
  {Phys. Rev. Lett.}\ }\textbf {\bibinfo {volume} {112}},\ \bibinfo {pages}
  {013002} (\bibinfo {year} {2014})}\BibitemShut {NoStop}%
\bibitem [{\citenamefont {Lesanovsky}\ and\ \citenamefont
  {Garrahan}(2013)}]{PRL-KinC}%
  \BibitemOpen
  \bibfield  {author} {\bibinfo {author} {\bibfnamefont {I.}~\bibnamefont
  {Lesanovsky}}\ and\ \bibinfo {author} {\bibfnamefont {J.~P.}\ \bibnamefont
  {Garrahan}},\ }\href@noop {} {\bibfield  {journal} {\bibinfo  {journal}
  {Phys. Rev. Lett.}\ }\textbf {\bibinfo {volume} {111}},\ \bibinfo {pages}
  {215305} (\bibinfo {year} {2013})}\BibitemShut {NoStop}%
\bibitem [{\citenamefont {Lesanovsky}\ and\ \citenamefont
  {Garrahan}(2014)}]{Lesanovsky2014}%
  \BibitemOpen
  \bibfield  {author} {\bibinfo {author} {\bibfnamefont {I.}~\bibnamefont
  {Lesanovsky}}\ and\ \bibinfo {author} {\bibfnamefont {J.~P.}\ \bibnamefont
  {Garrahan}},\ }\href {\doibase 10.1103/PhysRevA.90.011603} {\bibfield
  {journal} {\bibinfo  {journal} {Phys. Rev. A}\ }\textbf {\bibinfo {volume}
  {90}},\ \bibinfo {pages} {011603} (\bibinfo {year} {2014})}\BibitemShut
  {NoStop}%
\bibitem [{\citenamefont {Urvoy}\ \emph {et~al.}(2015)\citenamefont {Urvoy},
  \citenamefont {Ripka}, \citenamefont {Lesanovsky}, \citenamefont {Booth},
  \citenamefont {Shaffer}, \citenamefont {Pfau},\ and\ \citenamefont
  {L\"ow}}]{Urvoy2015}%
  \BibitemOpen
  \bibfield  {author} {\bibinfo {author} {\bibfnamefont {A.}~\bibnamefont
  {Urvoy}}, \bibinfo {author} {\bibfnamefont {F.}~\bibnamefont {Ripka}},
  \bibinfo {author} {\bibfnamefont {I.}~\bibnamefont {Lesanovsky}}, \bibinfo
  {author} {\bibfnamefont {D.}~\bibnamefont {Booth}}, \bibinfo {author}
  {\bibfnamefont {J.~P.}\ \bibnamefont {Shaffer}}, \bibinfo {author}
  {\bibfnamefont {T.}~\bibnamefont {Pfau}}, \ and\ \bibinfo {author}
  {\bibfnamefont {R.}~\bibnamefont {L\"ow}},\ }\href {\doibase
  10.1103/PhysRevLett.114.203002} {\bibfield  {journal} {\bibinfo  {journal}
  {Phys. Rev. Lett.}\ }\textbf {\bibinfo {volume} {114}},\ \bibinfo {pages}
  {203002} (\bibinfo {year} {2015})}\BibitemShut {NoStop}%
\bibitem [{\citenamefont {{Mattioli}}\ \emph {et~al.}(2015)\citenamefont
  {{Mattioli}}, \citenamefont {{Glaetzle}},\ and\ \citenamefont
  {{Lechner}}}]{Mattioli2015}%
  \BibitemOpen
  \bibfield  {author} {\bibinfo {author} {\bibfnamefont {M.}~\bibnamefont
  {{Mattioli}}}, \bibinfo {author} {\bibfnamefont {A.~W.}\ \bibnamefont
  {{Glaetzle}}}, \ and\ \bibinfo {author} {\bibfnamefont {W.}~\bibnamefont
  {{Lechner}}},\ }\href@noop {} {\bibfield  {journal} {\bibinfo  {journal}
  {ArXiv e-prints}\ } (\bibinfo {year} {2015})},\ \Eprint
  {http://arxiv.org/abs/1506.00906} {arXiv:1506.00906 [quant-ph]} \BibitemShut
  {NoStop}%
\bibitem [{\citenamefont {Hoening}\ \emph {et~al.}(2014)\citenamefont
  {Hoening}, \citenamefont {Abdussalam}, \citenamefont {Fleischhauer},\ and\
  \citenamefont {Pohl}}]{AF-num1}%
  \BibitemOpen
  \bibfield  {author} {\bibinfo {author} {\bibfnamefont {M.}~\bibnamefont
  {Hoening}}, \bibinfo {author} {\bibfnamefont {W.}~\bibnamefont {Abdussalam}},
  \bibinfo {author} {\bibfnamefont {M.}~\bibnamefont {Fleischhauer}}, \ and\
  \bibinfo {author} {\bibfnamefont {T.}~\bibnamefont {Pohl}},\ }\href {\doibase
  10.1103/PhysRevA.90.021603} {\bibfield  {journal} {\bibinfo  {journal} {Phys.
  Rev. A}\ }\textbf {\bibinfo {volume} {90}},\ \bibinfo {pages} {021603}
  (\bibinfo {year} {2014})}\BibitemShut {NoStop}%
\bibitem [{\citenamefont {H\"oning}\ \emph {et~al.}(2013)\citenamefont
  {H\"oning}, \citenamefont {Muth}, \citenamefont {Petrosyan},\ and\
  \citenamefont {Fleischhauer}}]{tDMRG1}%
  \BibitemOpen
  \bibfield  {author} {\bibinfo {author} {\bibfnamefont {M.}~\bibnamefont
  {H\"oning}}, \bibinfo {author} {\bibfnamefont {D.}~\bibnamefont {Muth}},
  \bibinfo {author} {\bibfnamefont {D.}~\bibnamefont {Petrosyan}}, \ and\
  \bibinfo {author} {\bibfnamefont {M.}~\bibnamefont {Fleischhauer}},\ }\href
  {\doibase 10.1103/PhysRevA.87.023401} {\bibfield  {journal} {\bibinfo
  {journal} {Phys. Rev. A}\ }\textbf {\bibinfo {volume} {87}},\ \bibinfo
  {pages} {023401} (\bibinfo {year} {2013})}\BibitemShut {NoStop}%
\bibitem [{\citenamefont {Lee}\ \emph {et~al.}(2011)\citenamefont {Lee},
  \citenamefont {H\"affner},\ and\ \citenamefont {Cross}}]{FullDynMF}%
  \BibitemOpen
  \bibfield  {author} {\bibinfo {author} {\bibfnamefont {T.~E.}\ \bibnamefont
  {Lee}}, \bibinfo {author} {\bibfnamefont {H.}~\bibnamefont {H\"affner}}, \
  and\ \bibinfo {author} {\bibfnamefont {M.~C.}\ \bibnamefont {Cross}},\ }\href
  {\doibase 10.1103/PhysRevA.84.031402} {\bibfield  {journal} {\bibinfo
  {journal} {Phys. Rev. A}\ }\textbf {\bibinfo {volume} {84}},\ \bibinfo
  {pages} {031402} (\bibinfo {year} {2011})}\BibitemShut {NoStop}%
\bibitem [{\citenamefont {Ates}\ \emph
  {et~al.}(2012{\natexlab{b}})\citenamefont {Ates}, \citenamefont {Olmos},
  \citenamefont {Garrahan},\ and\ \citenamefont {Lesanovsky}}]{PRA-Int}%
  \BibitemOpen
  \bibfield  {author} {\bibinfo {author} {\bibfnamefont {C.}~\bibnamefont
  {Ates}}, \bibinfo {author} {\bibfnamefont {B.}~\bibnamefont {Olmos}},
  \bibinfo {author} {\bibfnamefont {J.~P.}\ \bibnamefont {Garrahan}}, \ and\
  \bibinfo {author} {\bibfnamefont {I.}~\bibnamefont {Lesanovsky}},\ }\href
  {\doibase 10.1103/PhysRevA.85.043620} {\bibfield  {journal} {\bibinfo
  {journal} {Phys. Rev. A}\ }\textbf {\bibinfo {volume} {85}},\ \bibinfo
  {pages} {043620} (\bibinfo {year} {2012}{\natexlab{b}})}\BibitemShut
  {NoStop}%
\bibitem [{\citenamefont {Lee}\ \emph {et~al.}(2012)\citenamefont {Lee},
  \citenamefont {H\"affner},\ and\ \citenamefont {Cross}}]{Ryd-bistab1}%
  \BibitemOpen
  \bibfield  {author} {\bibinfo {author} {\bibfnamefont {T.~E.}\ \bibnamefont
  {Lee}}, \bibinfo {author} {\bibfnamefont {H.}~\bibnamefont {H\"affner}}, \
  and\ \bibinfo {author} {\bibfnamefont {M.~C.}\ \bibnamefont {Cross}},\ }\href
  {\doibase 10.1103/PhysRevLett.108.023602} {\bibfield  {journal} {\bibinfo
  {journal} {Phys. Rev. Lett.}\ }\textbf {\bibinfo {volume} {108}},\ \bibinfo
  {pages} {023602} (\bibinfo {year} {2012})}\BibitemShut {NoStop}%
\bibitem [{\citenamefont {Carr}\ \emph {et~al.}(2013)\citenamefont {Carr},
  \citenamefont {Ritter}, \citenamefont {Wade}, \citenamefont {Adams},\ and\
  \citenamefont {Weatherill}}]{Experiment1}%
  \BibitemOpen
  \bibfield  {author} {\bibinfo {author} {\bibfnamefont {C.}~\bibnamefont
  {Carr}}, \bibinfo {author} {\bibfnamefont {R.}~\bibnamefont {Ritter}},
  \bibinfo {author} {\bibfnamefont {C.~G.}\ \bibnamefont {Wade}}, \bibinfo
  {author} {\bibfnamefont {C.~S.}\ \bibnamefont {Adams}}, \ and\ \bibinfo
  {author} {\bibfnamefont {K.~J.}\ \bibnamefont {Weatherill}},\ }\href@noop {}
  {\bibfield  {journal} {\bibinfo  {journal} {Phys. Rev. Lett.}\ }\textbf
  {\bibinfo {volume} {111}},\ \bibinfo {pages} {113901} (\bibinfo {year}
  {2013})}\BibitemShut {NoStop}%
\bibitem [{\citenamefont {Marcuzzi}\ \emph
  {et~al.}(2014{\natexlab{a}})\citenamefont {Marcuzzi}, \citenamefont {Levi},
  \citenamefont {Diehl}, \citenamefont {Garrahan},\ and\ \citenamefont
  {Lesanovsky}}]{MM2014}%
  \BibitemOpen
  \bibfield  {author} {\bibinfo {author} {\bibfnamefont {M.}~\bibnamefont
  {Marcuzzi}}, \bibinfo {author} {\bibfnamefont {E.}~\bibnamefont {Levi}},
  \bibinfo {author} {\bibfnamefont {S.}~\bibnamefont {Diehl}}, \bibinfo
  {author} {\bibfnamefont {J.~P.}\ \bibnamefont {Garrahan}}, \ and\ \bibinfo
  {author} {\bibfnamefont {I.}~\bibnamefont {Lesanovsky}},\ }\href {\doibase
  10.1103/PhysRevLett.113.210401} {\bibfield  {journal} {\bibinfo  {journal}
  {Phys. Rev. Lett.}\ }\textbf {\bibinfo {volume} {113}},\ \bibinfo {pages}
  {210401} (\bibinfo {year} {2014}{\natexlab{a}})}\BibitemShut {NoStop}%
\bibitem [{\citenamefont {Weimer}(2015)}]{Weimer2014}%
  \BibitemOpen
  \bibfield  {author} {\bibinfo {author} {\bibfnamefont {H.}~\bibnamefont
  {Weimer}},\ }\href {\doibase 10.1103/PhysRevLett.114.040402} {\bibfield
  {journal} {\bibinfo  {journal} {Phys. Rev. Lett.}\ }\textbf {\bibinfo
  {volume} {114}},\ \bibinfo {pages} {040402} (\bibinfo {year}
  {2015})}\BibitemShut {NoStop}%
\bibitem [{\citenamefont {{Weimer}}(2015)}]{Weimer2015}%
  \BibitemOpen
  \bibfield  {author} {\bibinfo {author} {\bibfnamefont {H.}~\bibnamefont
  {{Weimer}}},\ }\href {\doibase 10.1103/PhysRevA.91.063401} {\bibfield
  {journal} {\bibinfo  {journal} {\pra}\ }\textbf {\bibinfo {volume} {91}},\
  \bibinfo {eid} {063401} (\bibinfo {year} {2015})},\ \Eprint
  {http://arxiv.org/abs/1501.07284} {arXiv:1501.07284 [cond-mat.quant-gas]}
  \BibitemShut {NoStop}%
\bibitem [{\citenamefont {Malossi}\ \emph {et~al.}(2014)\citenamefont
  {Malossi}, \citenamefont {Valado}, \citenamefont {Scotto}, \citenamefont
  {Huillery}, \citenamefont {Pillet}, \citenamefont {Ciampini}, \citenamefont
  {Arimondo},\ and\ \citenamefont {Morsch}}]{Experiment2}%
  \BibitemOpen
  \bibfield  {author} {\bibinfo {author} {\bibfnamefont {N.}~\bibnamefont
  {Malossi}}, \bibinfo {author} {\bibfnamefont {M.~M.}\ \bibnamefont {Valado}},
  \bibinfo {author} {\bibfnamefont {S.}~\bibnamefont {Scotto}}, \bibinfo
  {author} {\bibfnamefont {P.}~\bibnamefont {Huillery}}, \bibinfo {author}
  {\bibfnamefont {P.}~\bibnamefont {Pillet}}, \bibinfo {author} {\bibfnamefont
  {D.}~\bibnamefont {Ciampini}}, \bibinfo {author} {\bibfnamefont
  {E.}~\bibnamefont {Arimondo}}, \ and\ \bibinfo {author} {\bibfnamefont
  {O.}~\bibnamefont {Morsch}},\ }\href {\doibase
  10.1103/PhysRevLett.113.023006} {\bibfield  {journal} {\bibinfo  {journal}
  {Phys. Rev. Lett.}\ }\textbf {\bibinfo {volume} {113}},\ \bibinfo {pages}
  {023006} (\bibinfo {year} {2014})}\BibitemShut {NoStop}%
\bibitem [{\citenamefont {Marcuzzi}\ \emph
  {et~al.}(2014{\natexlab{b}})\citenamefont {Marcuzzi}, \citenamefont {Levi},
  \citenamefont {Li}, \citenamefont {Garrahan}, \citenamefont {Olmos},\ and\
  \citenamefont {Lesanovsky}}]{MM_DP}%
  \BibitemOpen
  \bibfield  {author} {\bibinfo {author} {\bibfnamefont {M.}~\bibnamefont
  {Marcuzzi}}, \bibinfo {author} {\bibfnamefont {E.}~\bibnamefont {Levi}},
  \bibinfo {author} {\bibfnamefont {W.}~\bibnamefont {Li}}, \bibinfo {author}
  {\bibfnamefont {J.~P.}\ \bibnamefont {Garrahan}}, \bibinfo {author}
  {\bibfnamefont {B.}~\bibnamefont {Olmos}}, \ and\ \bibinfo {author}
  {\bibfnamefont {I.}~\bibnamefont {Lesanovsky}},\ }\href@noop {} {\bibfield
  {journal} {\bibinfo  {journal} {arXiv:cond-mat/1411.7984}\ } (\bibinfo {year}
  {2014}{\natexlab{b}})}\BibitemShut {NoStop}%
\bibitem [{\citenamefont {Nogrette}\ \emph {et~al.}(2014)\citenamefont
  {Nogrette}, \citenamefont {Labuhn}, \citenamefont {Ravets}, \citenamefont
  {Barredo}, \citenamefont {B\'eguin}, \citenamefont {Vernier}, \citenamefont
  {Lahaye},\ and\ \citenamefont {Browaeys}}]{Nogrette2014}%
  \BibitemOpen
  \bibfield  {author} {\bibinfo {author} {\bibfnamefont {F.}~\bibnamefont
  {Nogrette}}, \bibinfo {author} {\bibfnamefont {H.}~\bibnamefont {Labuhn}},
  \bibinfo {author} {\bibfnamefont {S.}~\bibnamefont {Ravets}}, \bibinfo
  {author} {\bibfnamefont {D.}~\bibnamefont {Barredo}}, \bibinfo {author}
  {\bibfnamefont {L.}~\bibnamefont {B\'eguin}}, \bibinfo {author}
  {\bibfnamefont {A.}~\bibnamefont {Vernier}}, \bibinfo {author} {\bibfnamefont
  {T.}~\bibnamefont {Lahaye}}, \ and\ \bibinfo {author} {\bibfnamefont
  {A.}~\bibnamefont {Browaeys}},\ }\href {\doibase 10.1103/PhysRevX.4.021034}
  {\bibfield  {journal} {\bibinfo  {journal} {Phys. Rev. X}\ }\textbf {\bibinfo
  {volume} {4}},\ \bibinfo {pages} {021034} (\bibinfo {year}
  {2014})}\BibitemShut {NoStop}%
\bibitem [{\citenamefont {Wilk}\ \emph {et~al.}(2010)\citenamefont {Wilk},
  \citenamefont {Ga\"etan}, \citenamefont {Evellin}, \citenamefont {Wolters},
  \citenamefont {Miroshnychenko}, \citenamefont {Grangier},\ and\ \citenamefont
  {Browaeys}}]{Wilk2010}%
  \BibitemOpen
  \bibfield  {author} {\bibinfo {author} {\bibfnamefont {T.}~\bibnamefont
  {Wilk}}, \bibinfo {author} {\bibfnamefont {A.}~\bibnamefont {Ga\"etan}},
  \bibinfo {author} {\bibfnamefont {C.}~\bibnamefont {Evellin}}, \bibinfo
  {author} {\bibfnamefont {J.}~\bibnamefont {Wolters}}, \bibinfo {author}
  {\bibfnamefont {Y.}~\bibnamefont {Miroshnychenko}}, \bibinfo {author}
  {\bibfnamefont {P.}~\bibnamefont {Grangier}}, \ and\ \bibinfo {author}
  {\bibfnamefont {A.}~\bibnamefont {Browaeys}},\ }\href {\doibase
  10.1103/PhysRevLett.104.010502} {\bibfield  {journal} {\bibinfo  {journal}
  {Phys. Rev. Lett.}\ }\textbf {\bibinfo {volume} {104}},\ \bibinfo {pages}
  {010502} (\bibinfo {year} {2010})}\BibitemShut {NoStop}%
\bibitem [{\citenamefont {Barredo}\ \emph {et~al.}(2015)\citenamefont
  {Barredo}, \citenamefont {Labuhn}, \citenamefont {Ravets}, \citenamefont
  {Lahaye}, \citenamefont {Browaeys},\ and\ \citenamefont
  {Adams}}]{Barredo2015}%
  \BibitemOpen
  \bibfield  {author} {\bibinfo {author} {\bibfnamefont {D.}~\bibnamefont
  {Barredo}}, \bibinfo {author} {\bibfnamefont {H.}~\bibnamefont {Labuhn}},
  \bibinfo {author} {\bibfnamefont {S.}~\bibnamefont {Ravets}}, \bibinfo
  {author} {\bibfnamefont {T.}~\bibnamefont {Lahaye}}, \bibinfo {author}
  {\bibfnamefont {A.}~\bibnamefont {Browaeys}}, \ and\ \bibinfo {author}
  {\bibfnamefont {C.~S.}\ \bibnamefont {Adams}},\ }\href {\doibase
  10.1103/PhysRevLett.114.113002} {\bibfield  {journal} {\bibinfo  {journal}
  {Phys. Rev. Lett.}\ }\textbf {\bibinfo {volume} {114}},\ \bibinfo {pages}
  {113002} (\bibinfo {year} {2015})}\BibitemShut {NoStop}%
\bibitem [{\citenamefont {Gaetan}\ \emph {et~al.}(2009)\citenamefont {Gaetan},
  \citenamefont {Miroshnychenko}, \citenamefont {Wilk}, \citenamefont {Chotia},
  \citenamefont {Viteau}, \citenamefont {Comparat}, \citenamefont {Pillet},
  \citenamefont {Browaeys},\ and\ \citenamefont {Grangier}}]{Gaetan2009}%
  \BibitemOpen
  \bibfield  {author} {\bibinfo {author} {\bibfnamefont {A.}~\bibnamefont
  {Gaetan}}, \bibinfo {author} {\bibfnamefont {Y.}~\bibnamefont
  {Miroshnychenko}}, \bibinfo {author} {\bibfnamefont {T.}~\bibnamefont
  {Wilk}}, \bibinfo {author} {\bibfnamefont {A.}~\bibnamefont {Chotia}},
  \bibinfo {author} {\bibfnamefont {M.}~\bibnamefont {Viteau}}, \bibinfo
  {author} {\bibfnamefont {D.}~\bibnamefont {Comparat}}, \bibinfo {author}
  {\bibfnamefont {P.}~\bibnamefont {Pillet}}, \bibinfo {author} {\bibfnamefont
  {A.}~\bibnamefont {Browaeys}}, \ and\ \bibinfo {author} {\bibfnamefont
  {P.}~\bibnamefont {Grangier}},\ }\href {\doibase 10.1038/nphys1183}
  {\bibfield  {journal} {\bibinfo  {journal} {Nat. Phys.}\ }\textbf {\bibinfo
  {volume} {5}},\ \bibinfo {pages} {115} (\bibinfo {year} {2009})}\BibitemShut
  {NoStop}%
\bibitem [{\citenamefont {Marcuzzi}\ \emph
  {et~al.}(2014{\natexlab{c}})\citenamefont {Marcuzzi}, \citenamefont {Schick},
  \citenamefont {Olmos},\ and\ \citenamefont {Lesanovsky}}]{MM2014-2}%
  \BibitemOpen
  \bibfield  {author} {\bibinfo {author} {\bibfnamefont {M.}~\bibnamefont
  {Marcuzzi}}, \bibinfo {author} {\bibfnamefont {J.}~\bibnamefont {Schick}},
  \bibinfo {author} {\bibfnamefont {B.}~\bibnamefont {Olmos}}, \ and\ \bibinfo
  {author} {\bibfnamefont {I.}~\bibnamefont {Lesanovsky}},\ }\href
  {http://stacks.iop.org/1751-8121/47/i=48/a=482001} {\bibfield  {journal}
  {\bibinfo  {journal} {Journal of Physics A: Mathematical and Theoretical}\
  }\textbf {\bibinfo {volume} {47}},\ \bibinfo {pages} {482001} (\bibinfo
  {year} {2014}{\natexlab{c}})}\BibitemShut {NoStop}%
\bibitem [{\citenamefont {Schlosser}\ \emph {et~al.}(2001)\citenamefont
  {Schlosser}, \citenamefont {Reymond}, \citenamefont {Protsenko},\ and\
  \citenamefont {Grangier}}]{Schlosser2001}%
  \BibitemOpen
  \bibfield  {author} {\bibinfo {author} {\bibfnamefont {N.}~\bibnamefont
  {Schlosser}}, \bibinfo {author} {\bibfnamefont {G.}~\bibnamefont {Reymond}},
  \bibinfo {author} {\bibfnamefont {I.}~\bibnamefont {Protsenko}}, \ and\
  \bibinfo {author} {\bibfnamefont {P.}~\bibnamefont {Grangier}},\ }\href@noop
  {} {\bibfield  {journal} {\bibinfo  {journal} {Nature}\ }\textbf {\bibinfo
  {volume} {411}},\ \bibinfo {pages} {1024} (\bibinfo {year}
  {2001})}\BibitemShut {NoStop}%
\bibitem [{\citenamefont {Schlosser}\ \emph {et~al.}(2002)\citenamefont
  {Schlosser}, \citenamefont {Reymond},\ and\ \citenamefont
  {Grangier}}]{Schlosser2002}%
  \BibitemOpen
  \bibfield  {author} {\bibinfo {author} {\bibfnamefont {N.}~\bibnamefont
  {Schlosser}}, \bibinfo {author} {\bibfnamefont {G.}~\bibnamefont {Reymond}},
  \ and\ \bibinfo {author} {\bibfnamefont {P.}~\bibnamefont {Grangier}},\
  }\href {\doibase 10.1103/PhysRevLett.89.023005} {\bibfield  {journal}
  {\bibinfo  {journal} {Phys. Rev. Lett.}\ }\textbf {\bibinfo {volume} {89}},\
  \bibinfo {pages} {023005} (\bibinfo {year} {2002})}\BibitemShut {NoStop}%
\bibitem [{\citenamefont {Lukin}\ \emph {et~al.}(2001)\citenamefont {Lukin},
  \citenamefont {Fleischhauer}, \citenamefont {Cote}, \citenamefont {Duan},
  \citenamefont {Jaksch}, \citenamefont {Cirac},\ and\ \citenamefont
  {Zoller}}]{Lukin2001}%
  \BibitemOpen
  \bibfield  {author} {\bibinfo {author} {\bibfnamefont {M.}~\bibnamefont
  {Lukin}}, \bibinfo {author} {\bibfnamefont {M.}~\bibnamefont {Fleischhauer}},
  \bibinfo {author} {\bibfnamefont {R.}~\bibnamefont {Cote}}, \bibinfo {author}
  {\bibfnamefont {L.}~\bibnamefont {Duan}}, \bibinfo {author} {\bibfnamefont
  {D.}~\bibnamefont {Jaksch}}, \bibinfo {author} {\bibfnamefont
  {J.}~\bibnamefont {Cirac}}, \ and\ \bibinfo {author} {\bibfnamefont
  {P.}~\bibnamefont {Zoller}},\ }\href@noop {} {\bibfield  {journal} {\bibinfo
  {journal} {Physical Review Letters}\ }\textbf {\bibinfo {volume} {87}},\
  \bibinfo {pages} {037901} (\bibinfo {year} {2001})}\BibitemShut {NoStop}%
\bibitem [{\citenamefont {Urban}\ \emph {et~al.}(2009)\citenamefont {Urban},
  \citenamefont {Johnson}, \citenamefont {Henage}, \citenamefont {Isenhower},
  \citenamefont {Yavuz}, \citenamefont {Walker},\ and\ \citenamefont
  {Saffman}}]{Urban2009}%
  \BibitemOpen
  \bibfield  {author} {\bibinfo {author} {\bibfnamefont {E.}~\bibnamefont
  {Urban}}, \bibinfo {author} {\bibfnamefont {T.~A.}\ \bibnamefont {Johnson}},
  \bibinfo {author} {\bibfnamefont {T.}~\bibnamefont {Henage}}, \bibinfo
  {author} {\bibfnamefont {L.}~\bibnamefont {Isenhower}}, \bibinfo {author}
  {\bibfnamefont {D.}~\bibnamefont {Yavuz}}, \bibinfo {author} {\bibfnamefont
  {T.}~\bibnamefont {Walker}}, \ and\ \bibinfo {author} {\bibfnamefont
  {M.}~\bibnamefont {Saffman}},\ }\href@noop {} {\bibfield  {journal} {\bibinfo
   {journal} {Nature Physics}\ }\textbf {\bibinfo {volume} {5}},\ \bibinfo
  {pages} {110} (\bibinfo {year} {2009})}\BibitemShut {NoStop}%
\bibitem [{\citenamefont {Cummings}\ and\ \citenamefont
  {Dorri}(1983)}]{Cummings1983}%
  \BibitemOpen
  \bibfield  {author} {\bibinfo {author} {\bibfnamefont {F.~W.}\ \bibnamefont
  {Cummings}}\ and\ \bibinfo {author} {\bibfnamefont {A.}~\bibnamefont
  {Dorri}},\ }\href {\doibase 10.1103/PhysRevA.28.2282} {\bibfield  {journal}
  {\bibinfo  {journal} {Phys. Rev. A}\ }\textbf {\bibinfo {volume} {28}},\
  \bibinfo {pages} {2282} (\bibinfo {year} {1983})}\BibitemShut {NoStop}%
\bibitem [{\citenamefont {Raitzsch}\ \emph {et~al.}(2009)\citenamefont
  {Raitzsch}, \citenamefont {Heidemann}, \citenamefont {Weimer}, \citenamefont
  {Butscher}, \citenamefont {Kollmann}, \citenamefont {L{\"{o}}w},
  \citenamefont {B{\"u}chler},\ and\ \citenamefont {Pfau}}]{Dephasing}%
  \BibitemOpen
  \bibfield  {author} {\bibinfo {author} {\bibfnamefont {U.}~\bibnamefont
  {Raitzsch}}, \bibinfo {author} {\bibfnamefont {R.}~\bibnamefont {Heidemann}},
  \bibinfo {author} {\bibfnamefont {H.}~\bibnamefont {Weimer}}, \bibinfo
  {author} {\bibfnamefont {B.}~\bibnamefont {Butscher}}, \bibinfo {author}
  {\bibfnamefont {P.}~\bibnamefont {Kollmann}}, \bibinfo {author}
  {\bibfnamefont {R.}~\bibnamefont {L{\"{o}}w}}, \bibinfo {author}
  {\bibfnamefont {H.~P.}\ \bibnamefont {B{\"u}chler}}, \ and\ \bibinfo {author}
  {\bibfnamefont {T.}~\bibnamefont {Pfau}},\ }\href@noop {} {\bibfield
  {journal} {\bibinfo  {journal} {New J. Phys.}\ }\textbf {\bibinfo {volume}
  {11}},\ \bibinfo {pages} {055014} (\bibinfo {year} {2009})}\BibitemShut
  {NoStop}%
\bibitem [{\citenamefont {Lindblad}(1976)}]{Lindblad76}%
  \BibitemOpen
  \bibfield  {author} {\bibinfo {author} {\bibfnamefont {G.}~\bibnamefont
  {Lindblad}},\ }\href {\doibase 10.1007/BF01608499} {\bibfield  {journal}
  {\bibinfo  {journal} {Communications in Mathematical Physics}\ }\textbf
  {\bibinfo {volume} {48}},\ \bibinfo {pages} {119} (\bibinfo {year}
  {1976})}\BibitemShut {NoStop}%
\bibitem [{\citenamefont {Breuer}\ and\ \citenamefont
  {Petruccione}(2007)}]{Breuer_P}%
  \BibitemOpen
  \bibfield  {author} {\bibinfo {author} {\bibfnamefont {H.}~\bibnamefont
  {Breuer}}\ and\ \bibinfo {author} {\bibfnamefont {F.}~\bibnamefont
  {Petruccione}},\ }\href {https://books.google.co.uk/books?id=DkcJPwAACAAJ}
  {\emph {\bibinfo {title} {The Theory of Open Quantum Systems}}}\ (\bibinfo
  {publisher} {OUP Oxford},\ \bibinfo {year} {2007})\BibitemShut {NoStop}%
\bibitem [{\citenamefont {Ates}\ \emph {et~al.}(2006)\citenamefont {Ates},
  \citenamefont {Pohl}, \citenamefont {Pattard},\ and\ \citenamefont
  {Rost}}]{Ates06}%
  \BibitemOpen
  \bibfield  {author} {\bibinfo {author} {\bibfnamefont {C.}~\bibnamefont
  {Ates}}, \bibinfo {author} {\bibfnamefont {T.}~\bibnamefont {Pohl}}, \bibinfo
  {author} {\bibfnamefont {T.}~\bibnamefont {Pattard}}, \ and\ \bibinfo
  {author} {\bibfnamefont {J.~M.}\ \bibnamefont {Rost}},\ }\href {\doibase
  10.1088/0953-4075/39/11/L02} {\bibfield  {journal} {\bibinfo  {journal} {J.
  Phys. B}\ }\textbf {\bibinfo {volume} {39}},\ \bibinfo {pages} {L233}
  (\bibinfo {year} {2006})}\BibitemShut {NoStop}%
\bibitem [{\citenamefont {Heeg}\ \emph {et~al.}(2012)\citenamefont {Heeg},
  \citenamefont {G\"arttner},\ and\ \citenamefont {Evers}}]{Heeg2012}%
  \BibitemOpen
  \bibfield  {author} {\bibinfo {author} {\bibfnamefont {K.~P.}\ \bibnamefont
  {Heeg}}, \bibinfo {author} {\bibfnamefont {M.}~\bibnamefont {G\"arttner}}, \
  and\ \bibinfo {author} {\bibfnamefont {J.}~\bibnamefont {Evers}},\ }\href
  {\doibase 10.1103/PhysRevA.86.063421} {\bibfield  {journal} {\bibinfo
  {journal} {Phys. Rev. A}\ }\textbf {\bibinfo {volume} {86}},\ \bibinfo
  {pages} {063421} (\bibinfo {year} {2012})}\BibitemShut {NoStop}%
\bibitem [{\citenamefont {Nakajima}(1958)}]{Nakajima58}%
  \BibitemOpen
  \bibfield  {author} {\bibinfo {author} {\bibfnamefont {S.}~\bibnamefont
  {Nakajima}},\ }\href {\doibase 10.1143/PTP.20.948} {\bibfield  {journal}
  {\bibinfo  {journal} {Progr. Theor. Phys.}\ }\textbf {\bibinfo {volume}
  {20}},\ \bibinfo {pages} {948} (\bibinfo {year} {1958})}\BibitemShut
  {NoStop}%
\bibitem [{\citenamefont {Zwanzig}(1960)}]{Zwanzig60}%
  \BibitemOpen
  \bibfield  {author} {\bibinfo {author} {\bibfnamefont {R.}~\bibnamefont
  {Zwanzig}},\ }\href@noop {} {\bibfield  {journal} {\bibinfo  {journal} {J.
  Chem. Phys.}\ }\textbf {\bibinfo {volume} {33}},\ \bibinfo {pages} {1338}
  (\bibinfo {year} {1960})}\BibitemShut {NoStop}%
\bibitem [{\citenamefont {Degenfeld-Schonburg}\ and\ \citenamefont
  {Hartmann}(2014)}]{Degenfeld2014}%
  \BibitemOpen
  \bibfield  {author} {\bibinfo {author} {\bibfnamefont {P.}~\bibnamefont
  {Degenfeld-Schonburg}}\ and\ \bibinfo {author} {\bibfnamefont {M.~J.}\
  \bibnamefont {Hartmann}},\ }\href {\doibase 10.1103/PhysRevB.89.245108}
  {\bibfield  {journal} {\bibinfo  {journal} {Phys. Rev. B}\ }\textbf {\bibinfo
  {volume} {89}},\ \bibinfo {pages} {245108} (\bibinfo {year}
  {2014})}\BibitemShut {NoStop}%
\bibitem [{\citenamefont {Hinrichsen}(2000{\natexlab{a}})}]{DP_Hinrichsen}%
  \BibitemOpen
  \bibfield  {author} {\bibinfo {author} {\bibfnamefont {H.}~\bibnamefont
  {Hinrichsen}},\ }\href {\doibase 10.1080/00018730050198152} {\bibfield
  {journal} {\bibinfo  {journal} {Advances in Physics}\ }\textbf {\bibinfo
  {volume} {49}},\ \bibinfo {pages} {815} (\bibinfo {year}
  {2000}{\natexlab{a}})}\BibitemShut {NoStop}%
\bibitem [{\citenamefont {\'Odor}(2004)}]{Geza2004}%
  \BibitemOpen
  \bibfield  {author} {\bibinfo {author} {\bibfnamefont {G.}~\bibnamefont
  {\'Odor}},\ }\href {\doibase 10.1103/RevModPhys.76.663} {\bibfield  {journal}
  {\bibinfo  {journal} {Rev. Mod. Phys.}\ }\textbf {\bibinfo {volume} {76}},\
  \bibinfo {pages} {663} (\bibinfo {year} {2004})}\BibitemShut {NoStop}%
\bibitem [{\citenamefont {M.~Henkel}\ and\ \citenamefont
  {L{\"u}beck}(2009)}]{NEQ_PT1}%
  \BibitemOpen
  \bibfield  {author} {\bibinfo {author} {\bibfnamefont {H.~H.}\ \bibnamefont
  {M.~Henkel}}\ and\ \bibinfo {author} {\bibfnamefont {S.}~\bibnamefont
  {L{\"u}beck}},\ }\href@noop {} {\emph {\bibinfo {title} {Non-Equilibrium
  Phase Transitions}}},\ \bibinfo {series} {Theoretical and Mathematical
  Physics}, Vol.~\bibinfo {volume} {1}\ (\bibinfo  {publisher} {Springer},\
  \bibinfo {year} {2009})\BibitemShut {NoStop}%
\bibitem [{\citenamefont {Jensen}\ and\ \citenamefont
  {Dickman}(1993)}]{Jensen1993}%
  \BibitemOpen
  \bibfield  {author} {\bibinfo {author} {\bibfnamefont {I.}~\bibnamefont
  {Jensen}}\ and\ \bibinfo {author} {\bibfnamefont {R.}~\bibnamefont
  {Dickman}},\ }\href {\doibase 10.1103/PhysRevE.48.1710} {\bibfield  {journal}
  {\bibinfo  {journal} {Phys. Rev. E}\ }\textbf {\bibinfo {volume} {48}},\
  \bibinfo {pages} {1710} (\bibinfo {year} {1993})}\BibitemShut {NoStop}%
\bibitem [{\citenamefont {Petrosyan}\ and\ \citenamefont
  {M\o{}lmer}(2014)}]{Petrosyan14}%
  \BibitemOpen
  \bibfield  {author} {\bibinfo {author} {\bibfnamefont {D.}~\bibnamefont
  {Petrosyan}}\ and\ \bibinfo {author} {\bibfnamefont {K.}~\bibnamefont
  {M\o{}lmer}},\ }\href@noop {} {\bibfield  {journal} {\bibinfo  {journal}
  {Phys. Rev. Lett.}\ }\textbf {\bibinfo {volume} {113}},\ \bibinfo {pages}
  {123003} (\bibinfo {year} {2014})}\BibitemShut {NoStop}%
\bibitem [{\citenamefont {Kiffner}\ \emph
  {et~al.}(2013{\natexlab{a}})\citenamefont {Kiffner}, \citenamefont {Li},\
  and\ \citenamefont {Jaksch}}]{Kiffner13}%
  \BibitemOpen
  \bibfield  {author} {\bibinfo {author} {\bibfnamefont {M.}~\bibnamefont
  {Kiffner}}, \bibinfo {author} {\bibfnamefont {W.}~\bibnamefont {Li}}, \ and\
  \bibinfo {author} {\bibfnamefont {D.}~\bibnamefont {Jaksch}},\ }\href@noop {}
  {\bibfield  {journal} {\bibinfo  {journal} {Phys. Rev. Lett.}\ }\textbf
  {\bibinfo {volume} {111}},\ \bibinfo {pages} {233003} (\bibinfo {year}
  {2013}{\natexlab{a}})}\BibitemShut {NoStop}%
\bibitem [{\citenamefont {Kiffner}\ \emph
  {et~al.}(2013{\natexlab{b}})\citenamefont {Kiffner}, \citenamefont {Li},\
  and\ \citenamefont {Jaksch}}]{Kiffner13-1}%
  \BibitemOpen
  \bibfield  {author} {\bibinfo {author} {\bibfnamefont {M.}~\bibnamefont
  {Kiffner}}, \bibinfo {author} {\bibfnamefont {W.}~\bibnamefont {Li}}, \ and\
  \bibinfo {author} {\bibfnamefont {D.}~\bibnamefont {Jaksch}},\ }\href@noop {}
  {\bibfield  {journal} {\bibinfo  {journal} {Phys. Rev. Lett.}\ }\textbf
  {\bibinfo {volume} {110}},\ \bibinfo {pages} {170402} (\bibinfo {year}
  {2013}{\natexlab{b}})}\BibitemShut {NoStop}%
\bibitem [{\citenamefont {Strogatz}(1994)}]{Strogatz_book}%
  \BibitemOpen
  \bibfield  {author} {\bibinfo {author} {\bibfnamefont {S.}~\bibnamefont
  {Strogatz}},\ }\href {https://books.google.co.uk/books?id=FIYHiBLWCJMC}
  {\emph {\bibinfo {title} {Nonlinear Dynamics and Chaos: With Applications to
  Physics, Biology, Chemistry, and Engineering}}},\ Advanced book program\
  (\bibinfo  {publisher} {Westview Press},\ \bibinfo {year} {1994})\BibitemShut
  {NoStop}%
\bibitem [{\citenamefont {Hu}\ \emph {et~al.}(2013)\citenamefont {Hu},
  \citenamefont {Lee},\ and\ \citenamefont {Clark}}]{Largecorr}%
  \BibitemOpen
  \bibfield  {author} {\bibinfo {author} {\bibfnamefont {A.}~\bibnamefont
  {Hu}}, \bibinfo {author} {\bibfnamefont {T.~E.}\ \bibnamefont {Lee}}, \ and\
  \bibinfo {author} {\bibfnamefont {C.~W.}\ \bibnamefont {Clark}},\ }\href
  {\doibase 10.1103/PhysRevA.88.053627} {\bibfield  {journal} {\bibinfo
  {journal} {Phys. Rev. A}\ }\textbf {\bibinfo {volume} {88}},\ \bibinfo
  {pages} {053627} (\bibinfo {year} {2013})}\BibitemShut {NoStop}%
\bibitem [{\citenamefont {Jin}\ \emph {et~al.}(2013)\citenamefont {Jin},
  \citenamefont {Rossini}, \citenamefont {Fazio}, \citenamefont {Leib},\ and\
  \citenamefont {Hartmann}}]{cme1}%
  \BibitemOpen
  \bibfield  {author} {\bibinfo {author} {\bibfnamefont {J.}~\bibnamefont
  {Jin}}, \bibinfo {author} {\bibfnamefont {D.}~\bibnamefont {Rossini}},
  \bibinfo {author} {\bibfnamefont {R.}~\bibnamefont {Fazio}}, \bibinfo
  {author} {\bibfnamefont {M.}~\bibnamefont {Leib}}, \ and\ \bibinfo {author}
  {\bibfnamefont {M.~J.}\ \bibnamefont {Hartmann}},\ }\href {\doibase
  10.1103/PhysRevLett.110.163605} {\bibfield  {journal} {\bibinfo  {journal}
  {Phys. Rev. Lett.}\ }\textbf {\bibinfo {volume} {110}},\ \bibinfo {pages}
  {163605} (\bibinfo {year} {2013})}\BibitemShut {NoStop}%
\bibitem [{\citenamefont {Hohenberg}\ and\ \citenamefont
  {Halperin}(1977)}]{HH}%
  \BibitemOpen
  \bibfield  {author} {\bibinfo {author} {\bibfnamefont {P.~C.}\ \bibnamefont
  {Hohenberg}}\ and\ \bibinfo {author} {\bibfnamefont {B.~I.}\ \bibnamefont
  {Halperin}},\ }\href@noop {} {\bibfield  {journal} {\bibinfo  {journal} {Rev.
  Mod. Phys.}\ }\textbf {\bibinfo {volume} {49}},\ \bibinfo {pages} {435}
  (\bibinfo {year} {1977})}\BibitemShut {NoStop}%
\bibitem [{\citenamefont {Hinrichsen}(2000{\natexlab{b}})}]{Hinrichsen_exp}%
  \BibitemOpen
  \bibfield  {author} {\bibinfo {author} {\bibfnamefont {H.}~\bibnamefont
  {Hinrichsen}},\ }\href
  {http://www.scielo.br/scielo.php?script=sci_arttext&pid=S0103-97332000000100007&nrm=iso}
  {\bibfield  {journal} {\bibinfo  {journal} {{Brazilian Journal of Physics}}\
  }\textbf {\bibinfo {volume} {30}},\ \bibinfo {pages} {69 } (\bibinfo {year}
  {2000}{\natexlab{b}})}\BibitemShut {NoStop}%
\bibitem [{\citenamefont {Takeuchi}\ \emph {et~al.}(2007)\citenamefont
  {Takeuchi}, \citenamefont {Kuroda}, \citenamefont {Chat\'e},\ and\
  \citenamefont {Sano}}]{DP_exp}%
  \BibitemOpen
  \bibfield  {author} {\bibinfo {author} {\bibfnamefont {K.~A.}\ \bibnamefont
  {Takeuchi}}, \bibinfo {author} {\bibfnamefont {M.}~\bibnamefont {Kuroda}},
  \bibinfo {author} {\bibfnamefont {H.}~\bibnamefont {Chat\'e}}, \ and\
  \bibinfo {author} {\bibfnamefont {M.}~\bibnamefont {Sano}},\ }\href {\doibase
  10.1103/PhysRevLett.99.234503} {\bibfield  {journal} {\bibinfo  {journal}
  {Phys. Rev. Lett.}\ }\textbf {\bibinfo {volume} {99}},\ \bibinfo {pages}
  {234503} (\bibinfo {year} {2007})}\BibitemShut {NoStop}%
\bibitem [{\citenamefont {Takeuchi}\ \emph {et~al.}(2009)\citenamefont
  {Takeuchi}, \citenamefont {Kuroda}, \citenamefont {Chat\'e},\ and\
  \citenamefont {Sano}}]{DP_exp_long}%
  \BibitemOpen
  \bibfield  {author} {\bibinfo {author} {\bibfnamefont {K.~A.}\ \bibnamefont
  {Takeuchi}}, \bibinfo {author} {\bibfnamefont {M.}~\bibnamefont {Kuroda}},
  \bibinfo {author} {\bibfnamefont {H.}~\bibnamefont {Chat\'e}}, \ and\
  \bibinfo {author} {\bibfnamefont {M.}~\bibnamefont {Sano}},\ }\href {\doibase
  10.1103/PhysRevE.80.051116} {\bibfield  {journal} {\bibinfo  {journal} {Phys.
  Rev. E}\ }\textbf {\bibinfo {volume} {80}},\ \bibinfo {pages} {051116}
  (\bibinfo {year} {2009})}\BibitemShut {NoStop}%
\end{thebibliography}%

\end{document}